\pdfoutput=1
\documentclass[a4paper]{article}

\usepackage{ifthen,relsize,multicol,marvosym,float}
\usepackage{tabulary,multirow,lastpage,lineno,wrapfig,array}
\usepackage[pdftex]{graphicx}
\usepackage{epic,eepic,epsfig}
\usepackage{siunitx}
\usepackage[polutonikogreek,french,german,english]{babel}

\usepackage[T1]{fontenc}
\usepackage{titling}
\usepackage[centertags]{amsmath}
\usepackage{pbox,fancyhdr,textcomp,enumitem,ragged2e}
\usepackage{amssymb,amsfonts,oldgerm,mathrsfs,ipa}
\usepackage[text={6in,8.25in},centering]{geometry}

\newcommand{\afourlengths}{\geometry{a4paper,text={15cm,22.6cm},centering}}

\newcommand{\bibtocsec}{\addcontentsline{toc}{section}{\hspace*{-1.3em}\numberline{}References}}

\newcommand{\hyperrefpdf}{\usepackage[colorlinks=true,citecolor=red,linkcolor=blue,urlcolor=blue,pdftex]{hyperref}}

\newcommand{\skipline}[1][1]{\vspace*{#1\baselineskip}}

\newcommand{\noitemsep}{\setlength{\itemsep}{0in}}

\newcommand{\smidge}{\hspace*{.1em}}

\newcommand{\equalsdf}{=_{\mbox{\tiny df}}}

\usepackage{chicago}

\afourlengths

\predate{\skipline[-.75] \begin{center} \large}
\postdate{\end{center} \skipline[-.75]}

\hyperrefpdf

\title{Framework Confirmation by Newtonian Abduction\thanks{This paper
    is forthcoming in \emph{Synthese}, 2019
    (doi:\href{http://dx.doi.org/10.1007/s11229-019-02400-9}
    {10.1007/s11229-019-02400-9}).  I thank Bill Harper for many
    enjoyable and edifying discussions about Newton, scientific
    reasoning, and evidence, and for having written such a wonderful
    book (Harper 2011).  I also thank him for detailed comments on an
    earlier draft of the paper, including catching an error in \S3 in
    my discussion of Hall, Brown, and the precessions of Mercury and
    the Moon.  I thank Tom Pashby for comments on an earlier draft,
    including interesting suggestions for elaboration in future work.
    I thank an anonymous referee for detailed criticisms and
    questions.  I am grateful to Howard Stein as well, as always, for
    many illuminating and pleasurable conversations on all of these
    matters.  Work for this paper was funded by grant CU 338/1-1 from
    the Deutsche Forschungsgemeinschaft.}}

\author{Erik Curiel\thanks{\textbf{Author's address}: Munich Center
    for Mathematical Philosophy, Ludwig-Maximilians-Universit\"at;
    Black Hole Initiative, Harvard University; Smithsonian
    Astrophysical Observatory, Radio and Geoastronomy Division;
    \textbf{email}: \href{mailto:erik@strangebeautiful.com}
    {\texttt{erik@strangebeautiful.com}}}}

\date{}

\begin{document}
\thispagestyle{empty}
\maketitle


\begin{quote}
  \begin{center}
    \textbf{ABSTRACT}
  \end{center}  

  The analysis of theory-confirmation generally takes the deductive
  form: show that a theory in conjunction with physical data and
  auxiliary hypotheses yield a prediction about phenomena; verify the
  prediction; provide a quantitative measure of the degree of
  theory-confirmation this yields.  The issue of confirmation for an
  entire framework (\emph{e}.\emph{g}., Newtonian mechanics \emph{en{}
    bloc}, as opposed, say, to Newton's theory of gravitation) either
  does not arise, or is dismissed in so far as frameworks are thought
  not to be the kind of thing that admits scientific confirmation.  I
  argue that there is another form of scientific reasoning that has
  not received philosophical attention, what I call Newtonian
  abduction, that does provide confirmation for frameworks as a whole,
  and does so in two novel ways.  (In particular, Newtonian abduction
  is \emph{not} inference to the best explanation, but rather is
  closer to Peirce's original idea of abduction.)  I further argue
  that Newtonian abduction is at least as important a form of
  reasoning in science as standard deductive and inductive forms.  The
  form is beautifully summed up by Maxwell (1876): ``The true method
  of physical reasoning is to begin with the phenomena and to deduce
  the forces from them by a direct application of the equations of
  motion.''
\end{quote}

\tableofcontents

\skipline

\begin{quote}
  For the whole difficulty of [natural] philosophy seems to be to
  discover the forces of nature from the phenomena of motions and then
  to demonstrate the other phenomena from these forces.

  \skipline[-1]

  \begin{flushright}
    Newton \\
    \emph{Principia}, ``Preface''
  \end{flushright}
\end{quote}

\skipline

\section{Types of Reasoning in Science}
\label{sec:reas-sci}

In so far as one can identify something like a ``standard view'' in
the philosophical debates over a broad and vague topic like
confirmation, and in so far as one can make precise the distinction
between frameworks and theories, it is the standard view in the
philosophical community that frameworks are such things as not to
admit confirmation.  \citeN{friedman-dyns-reason-addend}, for
instance, in his influential discussion of frameworks and how what he
calls their relativized \emph{a priori} principles allow theory to
make substantive contact with experimental data, explicitly argues
that frameworks, based on the foundational roles they play in
evidentiary reasoning, cannot admit confirmation in any standard sense
on pain of circularity.  Based on his view of the role frameworks play
in grounding the meaning of terms in theories and the possibility of
applying theories to reason about phenomena, \citeN{carnap-eso} was
at pains to emphasize that the choice of framework is just that, a
choice, a practical action and not the assertion of a proposition
whose possible truth-value or doxastic support can be evaluated in
rationally principled ways; as such, the issue of confirmation, which
he took to be expressible by a proposition with determinate cognitive
content, whose truth value could be objectively evaluated, did not
arise with regard to frameworks.  \citeN{kuhn-struc-sci-rev} is
perhaps the \emph{locus classicus} of such a view, with his
extravagant claim of the fundamental and incorrigible irrationality in
the acceptance and use of a framework (or
``paradigm'').\footnote{Those who advocate some form of holism about
  knowledge will also hold that frameworks admit of confirmation.
  Quine, the canonical example of a holist, thought that frameworks
  admit empirical confirmation because he thought \emph{everything}
  does, even pure mathematics.  My reasons for claiming that
  frameworks admit confirmation differ from those of the holist: as
  will become clear in the paper's argument, I claim they do only in
  so far as the representations they allow us to construct make
  substantive, direct contact with the physical systems they purport
  to treat.  That is certainly not true of pure mathematics \emph{qua}
  mathematics.  It is also the case that some advocates of inference
  to the best explanation (IBE), such as \citeN{lipton-ibe}, claim
  that it allows for confirmation of frameworks.  See
  footnote~\ref{fn:ibe} below for a comparison of the ways that
  advocates of IBE{} claim that it supports confirmation of frameworks
  to the ways I argue Newtonian abduction does.}

In this paper, I argue to the contrary.  There is a form of scientific
reasoning that has not received philosophical attention, which I call
\emph{Newtonian abduction}, whose successful application provides
direct and strong confirmatory support for frameworks as a whole, not
just for theories that one may formulate in the context of the
framework.  The two forms of confirmation that accrue to it, what I
call \emph{structural confirmation} and \emph{modal confirmation},
have not to my knowledge been characterized in the literature, and are
of interest in themselves.

Above and beyond its role in confirmation of frameworks, moreover, I
claim this form of reasoning lies at the heart of many of the most
important advances in physics since the time of Newton.  A detailed
defense of that claim is beyond the scope of this paper, but I hope
that my explication of it and the examples I adduce and discuss will
go a long way towards making that case.  In any event, my discussion
will show that it is an important and fundamental form of reasoning in
physics that is logically, conceptually and methodologically distinct
from hypothetico-deductive reasoning, inductive reasoning, and IBE.
As such it deserves investigation by philosophers of science, and one
can reasonably expect such investigation to offer new illumination on
many important problems besides confirmation, such as the nature of
scientific explanation and understanding, the character of evidential
relations and the kinds of epistemic warrant evidence may provide for
theoretical claims, the way that theory and experiment make contact
with each other in general, a variety of inter-theoretic relations
such as reduction and emergence, and so on.

The standard form of confirmation most treated in the literature is,
in effect, that characterized by the following schema:
\begin{enumerate}
  \noitemsep
    \item theory + auxiliary hypotheses + initial conditions
  $\Rightarrow$ prediction;
    \item verify prediction; 
    \item apply qualitative or quantitative measure of degree of
  confirmation to theory.
\end{enumerate}
Of course, there are many substantive differences among the many
different approaches to the study of confirmation
\cite{earman-salmon-conf-sci-hypos,norton-little-survey-induc}, but
for my purposes using this schema as representative of the primary
ones suffices.  As the basis for philosophical reasoning about
confirmation, the schema is not wrong so far as it goes---but it goes
only so far.  It does not capture the confirmatory potential of other
forms of reasoning, and, most important for my purposes, not of
Newtonian abduction.\footnote{In order to try to avoid
  misunderstanding, I emphasize that I am not claiming that all forms
  of reasoning treated by systems, such as the Bayesian one, in which
  one tries to model confirmataory relations, are subsumed by the
  schema that I claim represents the ``standard form of
  confirmation''.  Philosophers have given strong arguments that the
  Bayesian framework, for example, can model the confirmatory
  relevance of forms of reasoning such as statistical inference,
  unification, explanation, and partial entailment.  I still think,
  nonetheless, that the---admittedly gross---simplification of using
  the logical form of HD as the comparative foil for the structure of
  confirmation in Newtonian abduction---as I will do throughout the
  paper---is justified for my purposes, since all the other forms of
  reasoning that, \emph{e}.\emph{g}., Bayesianism treats (unification,
  explanation, partial entailment, and so on) have the same difficulty
  capturing the kinds of confirmation that Newtonian abduction endows
  frameworks with, what I call below structural and modal
  confirmation.}

Roughly speaking, Newtonian abduction is the derivation of the
equations of motion, within a framework, of a particular type of
physical system, based on a given class of evolutions that type of
system may manifest: given a framework and a class of observed
physical evolutions of a type of system, one deduces from these the
unique theory of that type of system as formulated in the framework.
Newtonian abduction shows the framework to be capable of accounting
for phenomena by providing generic forms of equations of motion that,
in the framework's context, are necessary and sufficient for the sound
representation of the observed phenomena, and from which further
specific predictions can be derived.\footnote{\label{fn:eoms}I will
  speak of ``equations of motion'' in this paper, but it should be
  understood that what I say applies also to field equations, and
  indeed to any mathematical relations and physical principles a
  framework or theory posits as governing and constraining the
  behavior of the types of physical system it treats.  I will discuss
  examples of more general such types of relation and principle at the
  end of \S\ref{sec:abd}.}  It therefore, in a sense, when available,
\emph{grounds} the standard form of confirmation based on deductive
and other forms of reasoning, as adumbrated in the schema above.

Before going any further, I must emphasize that Newtonian abduction
has nothing to do with IBE, which is these days often referred to as
`abduction'.  Although the possibility of confusion is not negligible,
I like the name `Newtonian abduction' for the type of reasoning I
characterize for two reasons.  First, it was, to the best of my
knowledge, first and most consistently and powerfully deployed by
Newton.  Second, it corresponds closely in logical form---or perhaps
logical \emph{intent} is more to the point---to the idea of abduction
as originally promulgated by
\citeN{peirce-ded-ind-hypo,peirce-nat-mean,peirce-abd-ind}.\footnote{It
  seems to be the case that many if not most contemporary philosophers
  believe that Peirce's notion of abduction is in fact IBE.  That is
  wrong.  An attentive reading of Peirce shows that his notion of
  abduction was not equivalent to contemporary conceptions of IBE, and
  that neither did he ever champion any such idea.  Detailed exegesis
  to demonstrate this is beyond the scope of this paper.  I will only
  invite the reader to read
  \citeN{peirce-ded-ind-hypo,peirce-nat-mean,peirce-abd-ind} with
  an unbiased eye, and note that nowhere does Peirce use the
  ``goodness'' of the hypothesis as grounds for formulating it based
  on the evidence.  He certainly expected that one would, in the
  natural course of events, evaluate the goodness of the hypothesis
  once one had produced it by abduction, but the goodness of the
  hypothesis plays no role for him in its production by abductive
  reasoning.  As I will argue, the same holds true for Newtonian
  abduction.}  I will briefly explain at the end of \S\ref{sec:abd},
after I explicate and characterize the idea of Newtonian abduction,
this similarity of the form and intent, to justify my use of the term.
Also Newtonian abduction is not the same as what is known in the
literature as Newton's ``deduction from the phenomena''
\cite{harper-newt-class-deduc-pha,worrall-scope-lims-meth-deduc-pha},
though the two are related in important ways; indeed, as we will see,
one may with some justice view the standard account of Newton's
``deduction from the phenomena'' as capturing part of the full
structure of Newtonian abduction.

The rest of the paper is as follows.  In \S\ref{sec:fws}, I
characterize what I mean by a framework.  In \S\ref{sec:abd}, I
characterize Newtonian abduction as a form of reasoning within
frameworks, and discuss its salient logical and conceptual features.
In \S\ref{sec:fw-conf}, I sketch the way that successful Newtonian
abduction lends direct and immediate confirmatory support to
frameworks, and characterize the kinds of confirmation that accrue to
it.  I conclude in \S\ref{sec:true-confirm} by arguing against a
possible objection to the claim that Newtonian abduction lends
confirmatory support to the framework itself, and by discussing the
relation of Newtonian abduction to the idea of understanding in
science; I also briefly discuss the disconfirmation of frameworks.

\section{Frameworks}
\label{sec:fws}

In this section, I explain what I mean by `framework'.  I neither
promise nor threaten you with anything near a complete, precise
definition.  For my purposes, it suffices to provide only a rough
characterization sketched in broad strokes.\footnote{I discuss almost
  all the issues here in greater detail and to greater depth in
  \citeN{curiel-schem-obsr-epi-cont-theors}.}

A framework is a system that allows one to formulate propositions and
affirm them in principled ways based on evidence gathered according to
good principles and to apply these in turn as evidence based on good
principles.  So far, that does not differentiate it from a theory.  It
is so differentiated by the fact that some of the propositions it
allows one to formulate are themselves theories.\footnote{My
  conception of a framework is in many ways similar to, and indeed
  inspired by, Carnap's conception of a linguistic framework
  \cite{carnap-eso}, particularly in the way that a framework in
  both senses serves to define a fixed sense of physical possibility
  relevant to the kinds of system the framework treats.  Carnap's
  conception is too broad and vague, however, to do the work I require
  of it.  \citeN{stein-carnap-not-wrong} provides an insightful and
  illuminating, albeit brief, discussion of the differences between a
  Carnapian framework in the original sense and a framework in the
  sense of a structure in theoretical physics of the type I am
  sketching here, though he restricts attention to the strictly
  mathematical parts of one, as I do not.
  \citeN{lakatos-fals-meth-sci-rsrch} has some affinity with the gist
  of this view, in particular his notion of the ``hard core'' of
  ``research programs'', though, again, the differences in detail
  outweigh the similarities.  There is perhaps more affinity with the
  ``research traditions'' of \citeN{laudan-prog-probs}, in so far as
  different ones can share and swap important methodological and
  theoretical principles, as can happen with frameworks in my sense.
  A discussion of these comparisons is beyond the scope of this
  paper.}  A theory in this sense is a system that allows for the
unified representation and modeling of a particular kind of physical
system so as to render that kind amenable to investigation by
scientific reasoning and practices of all forms.  A particular kind of
physical system is one such that all individuals falling under the
kind bear the same physical quantities whose properties are
characterized by and whose behavior is governed by the same set of
equations of motion and collateral mathematical relations, such as
kinematical constraints \cite{curiel-kins-dyns-struc-theors}.  A
theory in my sense also includes such things as accounts of
experimental devices appropriate for the study of the relevant kind of
physical system, good practices for employing them, sound techniques
for the collection of raw data and statistical and other analysis and
organization of the same, reliable methods of approximative and
heuristic reasoning for constructing models and solving equations, and
guidelines for determining whether a system of the given kind is in a
state and experiencing interactions with its environment such as to be
jointly amenable to appropriate and adequate representation by the
theory (\emph{viz}., whether or not the system falls into the theory's
regime of applicability, as I will call it), and so on.\footnote{It is
  also characteristic of an appropriately unified kind of physical
  system, one treated by a theory in my sense, that there exist a set
  of scales at each of which all quantities the theory attributes to
  the kind of system simultaneously lose definition---the breakdown
  scales---as does not happen with the entirety of the family of all
  types of physical system treated by a framework.  In other words,
  every theory has a single, unified regime of applicability, bounded
  on all sides by scales characterized by the values of different
  combinations of its physical quantities.  For classical fluids, for
  example, the definitions of their pressure, fluid flow, viscosity,
  and all the rest break down at spatial and temporal scales a few
  orders of magnitude greater than those of the mean free-path of the
  fluid's constituent particles.  There is no \emph{a priori} reason
  why the definitions of all the different physical quantities
  represented by the theory should fail at the same characteristic
  scales, even though, in fact, those of all known theories do, not
  only for classical fluids but for all physical theories we have.
  This seems, indeed, to be one of the \emph{markers} of a physical
  theory, the existence of a set of characteristic scales for its
  physical quantities, at each of which all the theory's physical
  quantities simultaneously lose definition.  This is a fact that
  deserves philosophical investigation.}  These ideas are for my
purposes adequately clarified by examples.

Newtonian mechanics, the heart of which is embodied in the definitions
and the three laws Newton lays out in his \emph{Principia}, is a
framework.  Newton's Second Law in the abstract is not the equation of
motion of any particular kind of physical system.  It is rather the
schema that any equation of motion for any particular kind of system
treated by the framework must instantiate.  Newton's theory of gravity
is a theory formulated in the framework of Newtonian mechanics.  It
treats that kind of physical system characterized, \emph{inter alia},
by the possession of inertial mass, gravitational mass, spatial
position, and a velocity expressible as the temporal derivative of
spatial position, all such that the system's dynamical evolution is
governed by Newton's gravitational force law.  Navier-Stokes theory,
the classical theory of thermoconducive, visc\"oelastic fluids, is
another theory formulated in the framework of Newtonian mechanics.  It
treats that kind of physical system characterized, \emph{inter alia},
by the possession of inertial mass, shear viscosity, bulk viscosity,
thermal conductivity, fluid velocity, heat flux, pressure, and
shear-stress, all satisfying among themselves fixed relations of
constraint (\emph{e}.\emph{g}., that heat flux is always independent
of the pressure gradient), and whose dynamical evolution is governed
by the Navier-Stokes equations.\footnote{See \citeN[ch.~\textsc{v},
  \S49]{landau-lifschitz-fluid} for an exposition of the physics of
  Navier-Stokes theory, and \citeN{curiel-prpy-basis-sems} for a
  deeper and more extensive discussion of Navier-Stokes theory with
  regard to the issues I discuss here.  There is a subtlety I am
  glossing over: Navier-Stokes theory requires partial-differential
  equations for its formulation, which were not a part of the original
  Newtonian framework.  Strictly speaking, therefore, Navier-Stokes
  theory is formulated in an appropriate extension of the original
  Newtonian framework.  How frameworks can be extended in such ways is
  a fascinating problem, but one I cannot discuss here.}  One can also
formulate Navier-Stokes theory in the framework of Lagrangian
mechanics.  On the conception I put forward here, special relativity
is a framework, and one can formulate in it the theory of classical
Maxwell fields, as well as the theory of the mechanics of relativistic
point-particles.  Non-relativistic quantum mechanics is a framework,
within which one can formulate the atomic theory of elements of low
atomic weight.  Quantum field theory is a framework within which one
can formulate the theory of quantum electrodynamics.  Somewhat more
controversially, I suspect, according to my view general relativity is
a framework within which one can formulate theories of dust, fluids,
Maxwell fields, Dirac fields, and so on.\footnote{I argue elsewhere
  for the claim that general relativity is a framework in this sense,
  and not a theory.  Nothing in the paper hinges on the claim, so if
  you object to it, let it go.}

My analysis and arguments will need further distinctions with regard
to ``level'' of theoretical representation, so it will be useful to
characterize them by laying down terminology.  The schema in a
framework for equations of motion and other mathematical relations I
call \emph{abstract}.  Canonical examples are Newton's Second Law, the
Schr\"odinger equation in non-relativistic quantum mechanics, and so
on.  Structure and entities at the highest level of a theory
formulated in a given framework I will call \emph{generic}.  In
particular, generic structure has no definite values for those
quantities that appear as constants in the theory's equations of
motion and other mathematical relations.  The symbol `$k$' appearing
in the generic equation of motion of an elastic spring modeled as a
simple harmonic oscillator, $\ddot{x} = - \frac{k} {m} x$, denotes
Hooke's constant (the coefficient of proportionality between a force
applied to the spring and the resulting diplacement from its
equilibrium position), but possesses no fixed value, and the same for
the mass $m$.  It is important to keep in mind, however, that all
these formal representations of physical quantities do have
determinate physical dimensions, for Hooke's constant,
\emph{e}.\emph{g}.,
$\displaystyle \frac{\mathtt{mass}}{\mathtt{time}^\mathtt{2}}$.
Otherwise, one could not say that this is the generic equation of
motion for, say, a spring rather than a pendulum or oscillating string
or electric circuit or any other type of system whose dynamical
evolution is governed by the equation of motion for a simple harmonic
oscillator.

One can, in the same way, write down generic solutions to the generic
equations of motion.  These are formal representations of the
dynamically possible evolutions such systems can manifest.  One
generic solution to the equation of motion of an elastic spring is
$x(t) = \cos (\sqrt{\frac{k} {m}} \smidge t)$, where again one may
think of `$k$' and `$m$' as dummy variables, not determinate real
numbers.  Generic structure defines a \emph{genus} of physical system,
that type of physical system the theory appropriately and adequately
treats.

One obtains \emph{specific} structure by fixing the values of all such
constants in generic structure, say $m = 1$ and $k = 5$ (in some
system of units) for the elastic spring.  This defines a
\emph{species} of physical system of that genus, all springs with
those values for mass and Hooke's constant.\footnote{One can as well
  consider mixed systems, with, say, a fixed value for mass but
  indeterminate value for Hooke's constant.  These raise interesting
  questions, but they are beside the point here.}  One now has a
determinate space of states for systems of that species, and a
determinate family of dynamically possible evolutions, \emph{viz}.,
the solutions to the specific equations of motion, represented by a
distinguished family of paths on the space of states.  A path in this
family is an \emph{individual model} (or \emph{individual solution})
of the specific equations of motion; one common way of fixing an
individual model is by fixing definite initial conditions for the
specific equations of motion.  An individual model, as the name
suggests, represents a unique physical system of the species, that
whose dynamical quantities satisfy the initial conditions.  In the
case of the spring, that may mean the unique one whose position and
momentum at a given time have the values given by the initial
conditions.\footnote{I have deliberately taken my terminology from
  biological taxonomy, inspired by the remark of
  \citeN[p.~143]{peirce-doct-chnc}:
  \begin{quote}
    Now, the naturalists are the great builders of conceptions; there
    is no other branch of science where so much of this work is done
    as in theirs; and we must, to great measure, take them for our
    teachers in this important part of logic.
  \end{quote}
  There is much of insightful relevance in the lead-up to this remark,
  about how one individuates and characterizes genera and species of
  physical systems in my sense, which it would be illuminating to
  discuss, but it would take us too far afield.  I am tempted to
  describe structure at the level of a framework as \emph{phylar}, and
  to call the family of all types of physical system treated by a
  framework a \emph{phylum}---and so all physical systems would fall
  under the kingdom of physics---but I suspect it would just be
  distracting to the reader.}

Finally, I call a \emph{concrete model} a collection of experimentally
or observationally gathered results analyzed, structured and
interpreted in such a way as to allow identification with an
individual model; I will sometimes also refer to a concrete model as
\emph{structured data}.\footnote{This idea bears obvious and
  interesting comparison with the distinction between data and
  phenomena as drawn by \citeN{bogen-woodward-sav-pha}.  In so far as
  I understand their distinction, their idea of data more or less
  corresponds with my idea of ``experimentally or observationally
  gathered results'', but their notion of phenomena, in so far as it
  seems to try to capture something like general patterns in the
  world, does not neatly square with my conception of a concrete
  model, which is the result of appropriately transforming the results
  of a single experiment (or family of related experiments).}  If I
continually measure the position and momentum of a spring with known
values for $m$ and $k$, oscillating in one dimension for an interval
of time equal to its period of oscillation, and graph the results ``in
a natural way'', \emph{e}.\emph{g}., as a curve parametrized by time
on a Cartesian plane whose $x$-axis represents position and $y$-axis
represents momentum, then I will produce a curve that can be, ``in a
natural way'',\label{pg:sho-conc-mod} identified with exactly one
dynamically possible evolution on the space of states used to
represent that species of spring.  Kepler's organization and
structuring of the planetary ephemerides into parametrized ellipses
satisfying the Area Law and the Harmonic Law is an example that I will
rely on in some detail in the next section.  There are many
fascinating and deep problems associated with characterizing the ways
that experimental and observational results can be appropriately
transformed into structured data and identified with structures in
theory, but I put them aside for my purposes here.\footnote{I discuss
  many of them in detail in
  \citeN{curiel-schem-obsr-epi-cont-theors}.}

\section{Abduction}
\label{sec:abd}

I can now give a somewhat precise definition: \emph{Newtonian
  abduction} is the demonstration that a theory (generic equations of
motion) in a framework is logically equivalent to a concrete model of
a system appropriately treated by it.  It is the use of observed
dynamical evolutions of concrete physical systems to derive from an
abstract framework, such as Newtonian mechanics, the form of equations
of motion for systems of the given genus, such as Newton's
gravitational law.  It has the logical form
\begin{equation}
  \label{eq:log-form-nabd}
  \text{framework} \Rightarrow (\text{concrete model} \Leftrightarrow
  \text{generic equations of motion})
\end{equation}
where the arrows represent a relation of logical
entailment.\footnote{\label{fn:condl-defn}The abductive
  proposition~\eqref{eq:log-form-nabd} has the logical form of what
  \citeN[pp.~441\emph{ff}.]{carnap-test-mean-1} called a conditional
  definition.  I discuss the import of this below in
  \S\ref{sec:fw-conf}.}  All Newtonian abduction of the type of
primary interest here is of this particular form, involving a
framework, a concrete model, and generic equations of motion.  It
derives something (the theory) that is in a sense ``logically
intermediate'' between the other two terms in the formula, abstract
framework and concrete models, intermediate in the sense that one
standardly thinks of the framework as ``implying'' the generic
equations of motion, and then the generic equations of motion as
implying the individual models, which one then identifies with the
concrete models.  I will from hereon sometimes speak simply of
`abduction' or `abductive reasoning' and `abductive propositions';
unless explicitly stated otherwise, it should be understood that I
mean \emph{Newtonian}
abduction.\footnote{\label{fn:theory-med-meas}There are other types of
  Newtonian abduction, involving for example the derivation of
  specific equations of motion from generic equations of motion and
  concrete models.  One can, for instance, determine the ratio of
  Hooke's constant to the mass of a spring from a concrete model in
  conjunction with the generic equations of motion for a simple
  harmonic oscillator.  This is a simple-minded example of what
  \citeN{harper-newtons-sci-meth} and \citeN{smith-closing-loop} call
  \emph{theory-mediated measurement}.  More substantive examples are
  the determination of the relative masses of the planets in Newton's
  derivation of universal gravity and the determination of $\omega$ in
  Brans-Dicke theory by Shapiro time-delay
  \cite{harper-newtons-sci-meth}.  In these cases, the logical form
  of the reasoning is the same as in Newtonian abduction.  The
  difference is in the contents of the terms in the formula: the
  antecedent is the theory (\emph{e}.\emph{g}., Newtonian
  gravitational theory), not the framework; the lefthand side of the
  biconditional is still the concrete model; but the righthand side
  are the values of the parameters being determined.  I discuss this
  further in \S\ref{sec:fw-conf}.  One can also consider a weaker form
  of Newtonian abduction: not a biconditional with a single theory (a
  single set of generic equations of motion), but rather a
  biconditional with a family of related theories.  Later in this
  section, we will see an example of this in the discussion of
  Newton's framework for the investigation of light and color.}

Contrast this with induction, IBE{} and deduction.  A common form of
induction is: the construction of specific equations of motion as a
kind of generalization based on many concrete models.  A common form
of IBE{} is: the postulation of a set of specific equations of motion
based on argument to the effect that the equations of motion provide a
good explanation of the concrete models.  A common form of deduction
is: the derivation of an individual model from specific equations of
motion, auxiliary hypotheses, and initial conditions.  Its logical
form is
\begin{equation}
  \label{eq:log-form-hd}
  \text{(specific equations of motion)} \, \& \, \text{(initial
    conditions)} \, \& \, \text{(auxiliary hypotheses)} \Rightarrow
  \text{individual model}
\end{equation}
Indeed, this typifies what in philosophy standardly goes by the name
of hypothetico-deductive (HD) reasoning.  I will often use `HD' as
shorthand for all kinds of deductive reasoning and other forms of
inference (such as statistical) more or less conforming to this
schema.\footnote{I do not claim that all non-abductive forms of
  deductive reasoning and other similar forms of inference in science
  are HD; because details of difference in their form do not matter
  for my purposes, I ignore them.}  One of the most important
differences between this sort of HD reasoning and Newtonian abduction
is that the concrete model plays no role in the HD reasoning itself.
It rather comes into play only after the reasoning is complete, in the
attempt to identify the deduced individual model with a concrete model
in order to test the prediction.  The reason for this is simple: a
theory, in conjunction with initial data, can at most entail an
individual model; it cannot entail structured data derived from an
actual experiment.  In Newtonian abduction, to the contrary, the
identification of a concrete model with an individual model forms an
essential part of the reasoning itself that results in the theory.
This is possible because one has already the concrete model in hand,
as one does not in HD.  I discuss the consequences of this in more
detail in \S\ref{sec:fw-conf}.  An example will help illustrate the
differences among the forms of reasoning, as well as illuminating
perhaps the most important feature of Newtonian abduction,
\emph{viz}., that the derivation is a logical entailment of a
particular form, as with HD but not induction nor IBE.

In one of the most important steps in his argument in \emph{Principia}
for universal gravitation, Newton showed in Book~\textsc{iii} that the
magnitude of the force acting on the planets in their orbits about the
sun must be proportional to the squared inverse of the distance of the
planet from the sun, and that the direction along which the force acts
must lie along the line connecting the center of each planet and the
sun, pointing towards the sun.  To do this, he first proved in the
abstract framework of the system of dynamics he had developed, as
consequences of his Laws of Motion, that a system of bodies in
circular orbits about the same central body, whose orbits manifest a
certain set of fixed relations both individually and among themselves
(Kepler's Area and Harmonic Laws), must all experience a force obeying
an inverse-square law directed towards the central body.\footnote{The
  Area Law states that a planet in its orbit around the sun sweeps out
  equal areas in equal times, the area swept out being that of the
  region through which the line from the planet to the sun moves.  (In
  modern terms, this is equivalent to the conservation of angular
  momentum.)  The Harmonic Law states that, given the elliptical form
  of the orbits, the ratio of the square of the orbital period to the
  cube of the semi-major axis is the same for all planets.}  This was
a purely mathematical proposition, forming part of the abstract
structure of the framework.  He then noted, based on the concrete
models of the planetary orbits constructed \emph{\`a la} Kepler from
the best available astronomical observations of the day, that the
essentially circular orbits of the planets about the sun manifested
both individually and among themselves the relevant relations.  He
concluded that they experienced a force directed towards the sun,
obeying an inverse-square law.\footnote{The analysis of his arguments
  up to this step captures the heart of what is known in the
  literature as Newton's ``deduction from the phenomena''
  \cite{harper-newt-class-deduc-pha,worrall-scope-lims-meth-deduc-pha}.
  It is in this sense that the standard account of deduction from the
  phenomena captures part of Newtonian abduction.  As subsequent
  discussion will make clear, however, the standard account does not
  capture the full logical structure of Newtonian abduction, and thus
  cannot support the confirmatory weight I attribute to it.}  In other
words, using the resources of his abstract framework he abducted the
generic form of the system's equations of motion from their known
concrete dynamical evolutions.\footnote{There are several subtleties
  of the derivation I gloss over, such as his use of
  Corollaries~\textsc{vi} and \textsc{vii} to Proposition~\textsc{iv}
  in Book~\textsc{i}, which are propositions for concentric circles,
  not ellipses as he knew the planetary orbits to be.  The use of
  those propositions was justified because, when a planet in an
  elliptical orbit is at a distance equal to the semi-major axis
  (\emph{i}.\emph{e}., when it is 90 or 180 degrees from aphelion),
  then its centripetal acceleration exactly equals that of a planet in
  a uniform motion circular orbit of that radius and having the same
  period, and that suffices for the use of the corollaries.  See
  \citeN{stein-deduct-hypo} and
  \citeN[pp.~639\emph{ff}.]{stein-struct-know} for more detailed
  discussion of the logical structure of Newton's reasoning, and
  exposition and explanation of those sorts of subtleties.}

In the derivation, Newton did \emph{not} use Proposition~\textsc{i},
Book~\textsc{i}, which says that a centripetal force implies the Area
Law, to conclude that the force on the planets was always directed
along the line joining the center of the planet to that of the sun;
that would have been classic HD reasoning.  He rather uses
Proposition~\textsc{ii}, Book~\textsc{i}, which says that the Area Law
implies a centripetal force of the right kind.  He then invokes the
fact that the concrete planetary models satisfy Kepler's Harmonic Law
to deduce that the gravitational force between celestial bodies varies
as the inverse of the square of the distance between them, based on
Corollary~\textsc{vi} of Proposition~\textsc{iv}, Book~\textsc{i}.
Similarly, he does \emph{not} use the converse, that an inverse-square
force implies the Harmonic Law, which is a theorem as well (proven in
the same corollary), to argue for the inverse-square form of the
force.  Again, that would have been HD reasoning.  Finally, he did not
calculate, for each planet individually, that an inverse-square law
would produce the observed orbit, and then conclude based on these
instances that all relevantly similar bodies would also experience
such a force.  That would have been induction.

That the planets obey the Area and Harmonic Laws are propositions
derived from the Keplerian structuring of the direct astronomical
observation of the planets' positions over time; that fact, in
conjunction with the Newtonian framework (the Laws of Motion), implies
part of the form of the generic equations of motion (the force), that
it be centripetal, directed towards the Sun, and vary as the
inverse-square of the distance.  The argument that the force is
proportional to the product of the mass of each planet and the sun
follows similar reasoning.  Newton made use of Proposition~\textsc{i},
Book~\textsc{i}, that the centripetal form of the force implies the
Area Law, and that part of Corollary~\textsc{vi} to
Proposition~\textsc{iv}, Book~\textsc{i}, showing that the
inverse-square form of the force implies the Harmonic Law, only after
the derivation of universal gravity was complete, starting at
Proposition~\textsc{xiii}, Book~\textsc{iii}, with what he referred to
as an ``\emph{a priori}'' deduction of the motions of the celestial
bodies starting from the ``principles'' of his theory,
\emph{i}.\emph{e}., the generic equations of motion derived in the
abductive argument.  This explicitly shows that the logical form of
the relations among the framework, the concrete model, and the generic
equations of motion is that of Newtonian abduction: a conditional with
the framework as the antecedent, and a biconditional between the
concrete model and the generic equations of motion as the consequent,
conforming to formula~\eqref{eq:log-form-nabd}.  This is \emph{not}
ampliative reasoning in the sense that induction or IBE{} is.  It is
logical entailment.\footnote{It is important to note that, strictly
  speaking, Newton's initial ``deduction'' does not work, because, as
  Newton well knew, the concrete models he relied on did not exactly,
  only approximately, instantiate the Keplerian relations.  Here is
  where what \citeN{harper-newtons-sci-meth} calls Newton's method of
  successive approximations comes into play.  In effect, Newton
  decided to assert the validity of the simple abductive proposition
  and then push it as far as he could by successive refinements of the
  approximative models, each one markedly improving on the previous.
  (See \citeNP[pp.~177--180]{stein-newt-st} for a concise and lucid
  discussion of Newton's guiding methodological principle here and the
  way he applied it.)  It is worth remarking as well that HD cannot
  accommodate this method, as it can do nothing but strike the
  hypothesis from the record of viable candidates at the first glimpse
  of such inconsistency with the data.}

Newtonian abduction is thus the explicit integration of framework and
concrete experiment in such a way as to produce theories guaranteed to
be appropriate and adequate in all ways, including predictive
accuracy, for treating the phenomena used as the basis of the
abduction, at least up to the margin of error of the concrete
models.\footnote{It would be an interesting project to characterize
  the necessary and sufficient conditions on the structures
  (topological, algebraic, geometric, \emph{etc}.\@) a framework
  imposes on its family of theories required for the framework to
  support Newtonian abduction.}  In order to extend the scope of the
theory to show its propriety and adequacy for other physical systems
one has reason to think are of the same genus, one must perform
further reasoning, either of the abductive or HD type.  This is
exactly what Newton did in the remainder of Book~\textsc{iii}, using
abductive reasoning to show that comets and other celestial bodies
also obeyed the generic equations of motion of Newtonian gravitational
theory (following from the approximately parabolic form of their
orbits), and the same for terrestrial systems as well (free fall, the
motion of pendulums, and so on).

It should be clear that Newtonian abduction works as logical
entailment only because I have put into the framework from the start
all the practices and principles that allow one to meaningfully bring
theory and experiment into contact with each other so that the former
may be used to interpret the latter, and the latter may be used to
constrain the former, \emph{i}.\emph{e}., so that data may be
structured in such a way as to be directly comparable with, and even
identified with, solutions to the equations of motion
(\emph{i}.\emph{e}., the identification of concrete with individual
models).  This will include, \emph{inter alia}:
\begin{itemize}
  \noitemsep
    \item mathematical structure, relations and formul{\ae} in
  addition to the abstract equations of motion (\emph{e}.\emph{g}., in
  Newtonian mechanics that $\vec{v} \equalsdf \dot{\vec{x}}$, that
  mass is additive, that spacetime has a flat affine structure, and so
  on);
    \item standards of good argumentation for different genera of
  physical systems (accepted approximative techniques for solving
  equations and constructing simplified individual models, sound
  heuristics for informal arguments, and so on);
    \item families of accepted experimental and observational
  practices for systems of different genera;
    \item rules for connecting experimental outcomes with formal
  propositions (semantics, pragmatics, principles of representation,
  construction of concrete models from raw observations, rules for
  interpreting structured data using the conceptual vocabulary of the
  framework, rules for reckoning expected experimental precision and
  error, and so on);
    \item rules of evidential warrant (what can be evidence, how to
  apply it, reckoning of error tolerance, and so on);
    \item and guidelines for judging the legitimacy of proposed
  modifications, extensions, and restrictions of all these.  
\end{itemize}
In other words, the framework includes all physical and methodological
principles and practices required to bring theory and experiment (in
the form of structured data) into substantive, fruitful
contact.\footnote{On this view, the entirety of a framework is a
  dynamic entity, as are individual theories, evolving over time as
  new theoretical and experimental techniques and practices are
  developed and accepted, and so theories themselves will be as well.
  I think this is the right way to think about these matters for many
  if not most purposes in those parts of philosophy of science
  studying scientific theories.  The contemporary practice of treating
  theories as static, fixed entities, especially in work of a more
  technical and formal character, can lead to serious philosophical
  error.  An adequate semantics of a theory, for instance, should
  reflect and accommodate its dynamic nature.}  Otherwise the
entailment of the biconditional would not be valid.  Most of this will
be difficult if not impossible to articulate and record in an
exhaustive and precise way, so as to lend itself to use in formal
philosophical investigations.  We must trust that all such collateral
principles and practices are there, and can be, now and again, each
more or less precisely articulated as the occasion demands.  The same
holds true, however, for formal reconstructions of all forms of
reasoning in science (\emph{e}.\emph{g}., the `auxiliary hypotheses'
of HD, which always hide an ugly mob of philosophical sins).

Does all this make the idea of a framework too vague?  We simply throw
in everything that will ``make the abduction work''?  Is it,
\emph{e}.\emph{g}., a methodological principle of Newtonian mechanics
that one prefer a single law (inverse-square) for all the planets,
rather than, perhaps, $\displaystyle \frac{1} {r^{2.0000001612}}$ for
Mercury, the figure one arrives at by using the Precession Theorem
(explained below) on the most accurate data we have for Mercury's
orbit?\footnote{See \citeN{hall-sugg-theory-merc} for the initial
  proposal of this, and, \emph{e}.\emph{g}.,
  Newcomb~\citeyear{newcomb-4-inner-planets,newcomb-grav-encyc-brit}
  for further championing of the idea.}  To do the same to account for
the Moon's motion around the Earth, again based on the Precession
Theorem, would yield a force proportional to
$\displaystyle \frac{1} {r^{2 + \epsilon}}$, where
$\epsilon < 0.00000004$, clearly inconsistent with the exponent
required for Mercury, as shown by the analysis of
\citeN{brown-deg-acc-new-lun-thry}, of the Hill-Brown lunar theory.
(See \citeNP{wilson-newt-celest-mechs} for discussion.)  And one
would then go on to choose a slightly different force law for the
attraction of each individual planet to the sun, and for the
attraction of each comet to the sun, and so on.  All such laws could
be derived by abductive reasoning in the framework of Newtonian
mechanics.  Choosing those multiple force laws, of course, would
\emph{eo ipso} be to treat Newtonian gravitational theory as
disconfirmed.  But there \emph{are} methodological principles in the
Newtonian framework, explicitly in this case, that tell us to prefer
the single law for all planets, and indeed for all gravitating bodies,
\emph{viz}., Newton's Rules of Reasoning in Natural Philosophy
\cite[Book~\textsc{iii}]{newton-princ-motte}, specifically
Rules~\textsc{ii} and \textsc{iv}, which he invokes in the argument in
Book~\textsc{iii} for universal gravitation:
\begin{description}
    \item[Rule II] ``[T]o the same natural effects we must, as far as
  possible, assign the same causes.''
    \item[Rule IV] ``In experimental philosophy we are to look upon
  propositions inferred by general induction from phenomena as
  accurately or very nearly true, notwithstanding any contrary
  hypotheses that may be imagined, till such time as other phenomena
  occur, by which they may either be made more accurate, or liable to
  exceptions.''
\end{description}
The argument that these Rules imply, or at least militate in favor of,
a preference for a single force law is subtle and involved, and beyond
the scope of this paper, but I hope it to be at least intuitively
clear that they do so.\footnote{That Newton uses the term `general
  induction' in Rule \textsc{iv} has no bearing on my arguments.  The
  word `abduction' did not exist then, and, in any event, as is made
  clear by Newton's gloss on the rule following its statement, Newton
  is using `general induction' to refer, among other things, to the
  pattern of reasoning he employs in deriving his theory of universal
  gravity, which I claim is in fact Newtonian abduction.}

The Precession Theorem (\emph{Principia}, Book~\textsc{i},
Proposition~\textsc{xlv}, and its corollaries), and the use Newton and
later investigators put it to, display the great power of Newtonian
abduction as a form of reasoning, allowing one to draw conclusions
much stronger and more robust in several ways than any HD, inductive
or IBE{} reasoning can support.  The Precession Theorem, in essence,
says that, if a body is in nearly circular orbit around a central
body, and its perihelion precesses, then the rate of precession
measures the difference between an inverse-square force law and the
force law governing the orbit.  If such an orbit has no
precession---\emph{i}.\emph{e}., if it is closed---then it obeys an
exact inverse-square law.  If an orbit has small precession, then the
force law will be of the form
$\displaystyle \frac{1} {r^{2 + \epsilon}}$, for some small $\epsilon$
proportional to the amount of precession.  The converses hold as well:
if bodies do not obey an exact inverse-square force law, then there
will be precession, and its rate will measure the deviation from
inverse-square.  All the planetary orbits are, in the precise sense
Newton defines, nearly circular and exhibit small precession.  Based
on the best observational data, Newton knew that the orbit of no
planet is exactly a closed curve.  By calculating the expected
perturbative gravitational influences of the other planets (primarily
Jupiter) on the orbit of each planet, he showed that the precessions
would be negligible (on the order of arcminutes per century), and so
abductively concluded from the Precession Theorem that this provided
further evidence for the inverse-square law (Book~\textsc{iii},
Propositions~\textsc{xiii} and \textsc{xiv}).\footnote{More precisely,
  he used Proposition~\textsc{lxvi} and its corollaries
  (Book~\textsc{i}), which are themselves derived from the Precession
  Theorem.  The content of the Precession Theorem itself carries the
  burden of the argument.}  The only exceptions to this are the
perturbative effect Jupiter and Saturn have on each other when they
are close to conjunction, and the irregularities in the moon's motion
(\emph{e}.\emph{g}., librations arising from precession) caused by the
non-negligible combined effects on it of the oceans' tides and the
Sun.  Again, he was able to show by abductive argument based on the
Precession Theorem that more finely detailed concrete models of the
orbits, including the irregularities from perturbations, yield the
inverse-square force law (Book~\textsc{iii},
Propositions~\textsc{xiii}, \textsc{xvii}, and \textsc{xxii}).

\citeN{leverrier-mem-vars-orbs} subsequently improved on Newton's
reasoning.  Using more accurate data and more detailed perturbative
models of the effects of other planets' gravitation on a given
planet's orbit, he showed that subtracting the effective perturbative
forces would yield closed orbits for all the planets (within the
bounds of error for the available data), thus giving even stronger
confirmation of the inverse-square law by abduction based on the
Precession Theorem.\footnote{In fact, this is not true for Mercury, as
  Le Verrier well knew.  There was an extra 39 arcseconds ($39''$) per
  century of precession that his calculations could not account for.
  He labored for the next 14 years to produce a mechanism to explain
  the discrepancy, even postulating hitherto unobserved celestial
  bodies and other such \emph{ad hoc} devices, but nothing worked
  \cite{leverrier-merc}.  Indeed, by the end of the 19th Century the
  inexplicability of the aberrant precession was such a great
  embarrassment that many eminent physicists had already concluded
  that Newtonian gravitational theory could not be fundamentally
  correct, even before the development of special relativity (a
  historical fact that seems to be not so well known as it ought),
  \emph{based entirely on abductive use of the Precession Theorem}.
  See
  \citeN{newcomb-fund-consts-astro,newcomb-4-inner-planets,newcomb-precess-consts}
  for an extended discussion and summation of the experimental
  knowledge of the aberrant precession at that time, when the
  anomalous amount of Mercury's precession was finally fixed at $43''$
  per century, and see \citeN{freundlich-planet} for an exhaustive
  argument that Newtonian gravitational theory could not account for
  it.  To get a sense of how small the angle $43''$ is, imagine the
  appearance of the diameter of a penny from a distance of about 30
  miles.  The apparent length of its diameter on the eye, projected
  back to the penny, subtends an angle of that size.  It is a
  testament to the profound confirmatory entrenchment of Newtonian
  gravitational theory in particular at the time, and Newtonian
  mechanics in general, that a discrepancy of this infinitesimal angle
  \emph{per century} in a planetary orbit caused such consternation in
  and provoked such labor from the leading lights of the scientific
  community for more than 70 years.  Of course, we now know that the
  error arises from general relativistic effects, and cannot be
  accommodated by Newtonian gravitational theory.}  This is a powerful
improvement on Newton's initial argument I sketched above for the form
of the gravitational force law.  One constructs the initial individual
models used in deriving the inverse-square form, closed nearly
circular orbits, by ignoring known features of the concrete models
that deviate from the individual models, \emph{viz}., their
precession.  One then constructs new individual models that take into
account more of the known features of the concrete models, and
subsequently shows by abductive reasoning that those yet more accurate
individual models yield the inverse-square law, when corrective
perturbations are accounted for.

The form of the argument was not:
\begin{enumerate}
  \noitemsep
    \item if there are deviations from inverse-square, then there will
  be orbital precession (individual models);
    \item there is observed orbital precession (concrete models);
    \item therefore, I will find the right force law from which I can
  deduce the observed precession (individual models);
    \item I will do so by deducing all the expected perturbative
  effects of all other planets on each planet's orbit for different
  force laws individually (individual models);
    \item I will thus find the law that implies that those
  perturbative effects give rise to the observed precession.
\end{enumerate}
That would be HD reasoning.  Rather, the argument runs thus:
\begin{enumerate}
  \noitemsep
    \item there is observed orbital precession (concrete models);
    \item it is a mathematical theorem in the framework that there is
  orbital precession if and only if there is deviation from
  inverse-square, and the amount of precession determines the amount
  of deviation (abstract structure);
    \item I will construct individual models that include the
  perturbative effects from the other planets, based on the observed
  orbits (the concrete models);
    \item it then follows abductively, from the concrete models, the
  abstract equations of motion and the Precession Theorem, that there
  is a unique force law which is such that, for all planets, when the
  perturbations calculated from the force law are subtracted from each
  planet, and that planet is modeled as a two-body system orbiting
  around the sun, then the orbit of the two-body problem determined
  \emph{ab initio} by the force law is the same as the orbit
  determined by subtracting all the calculated perturbations, and that
  this holds if and only if the concrete model is uniquely
  identifiable with an appropriate individual model;
    \item the resulting orbits for all planets are closed,
  \emph{i}.\emph{e}., there is no precession;
    \item by abduction, this holds if and only if the unique force law
  is inverse-square.
\end{enumerate}

It is clear from the reconstruction, moreover, that abductive
reasoning supports reasoning based on subjunctive conditionals.  It
allows one to deduce conditionals of the form ``if the phenomena were
different in this particular way, the laws would be thus and so''.  In
fact, it allows one to deduce subjunctive biconditionals: ``the
phenomena would be different in this particular way if and only if the
laws were thus and so''.  It thus allows one to set principled bounds
on how far wrong the abductively derived laws can appear to be, and
still be within the known margins of error of extant observations.
Even more, it shows that construction of more finely detailed, more
accurate concrete models also yield the same force law by directly
showing, in one fell logical swoop, that no other force law
\emph{could} accurately represent the finer, more accurate models,
within the margin of error believed at the time.\footnote{Indeed,
  Newton had even more instances of such reasoning than only that
  based on the Precession Theorem.  According to
  Corollary~\textsc{vii} to Proposition~\textsc{iv}, Book~\textsc{i},
  the Harmonic Law also provides support for the derivation of such
  subjunctive conditionals.  Over and above the biconditional between
  the Harmonic Law and the inverse-square form of the force law, the
  corollary shows that the periods of the planetary orbits are
  proportional to a power greater than the Harmonic Law's 3/2 power of
  the semi-major axes if and only if the centripetal forces fall off
  more rapidly than the inverse squared law, and contrarily that the
  periods are proportional to a power less than 3/2 of the semi-major
  axes if and only if the centripetal forces fall off more slowly than
  the inverse-square law.  See \citeN{harper-1st-6-props-newt-arg}
  and \citeN[pp.~114--120]{harper-newtons-sci-meth} for discussion.}

Subjunctive reasoning of this kind provides extraordinarily strong
epistemic warrant for the abductively derived laws, as it shows how
far the concrete models constructed from detailed observations can
deviate from simpler individual models while still remaining within
the regime of applicability of the laws
\cite{harper-newtons-sci-meth,smith-closing-loop}.  This is not
possible in the HD scheme.  In that form of reasoning, one can show
the validity only of conditionals of the following form: ``if the laws
were different in this particular way, the phenomena would be thus and
so''.  But that shows nothing of epistemic import.  One cannot use
such conditionals to set bounds on how wrong simple individual models
can be and still fall within the regime of applicability of the
proposed laws when compared to more detailed concrete models.  Indeed,
one cannot use such conditionals to derive \emph{any substantive
  proposition at all} about the actual world.  This is a striking
example of the power of abduction as compared to standard
HD.\footnote{A good example of this from contemporary physics is the
  parametrized post-Einsteinian framework of
  \citeN{yunes-pretorius-theor-bias-grav-wv-ppe}, which manifestly
  supports subjunctive reasoning of this form, with the explicit
  intent of sharpening the recent observation of gravitational waves
  by LIGO \cite{abbott-et-obs-gw-bin-bh} as tests of general
  relativity, by providing a framework within which one can probe for
  deviations from general relativity's predictions in a parametrized,
  controllable form: the dynamics of the observed coalescence of the
  binary black hole system deviates from general relativity's models
  in this way, by this parametrized amount, if and only if the
  observed gravitational waveform exhibits this quantifiable feature.}

Newton's \citeyear{newton-light-color} framework for theories of
light and color makes a fascinating case study for teasing out the
many subtleties, layers and facets of these ideas.\footnote{For an
  exposition of Newton's framework for light and color, and the
  investigations and abductive reasoning that led him to it, see
  \citeN{curiel-modesty} (though I do not refer to the form of
  reasoning as Newtonian abduction in that paper), and for a more
  extensive and deeper discussion of those investigations, with a
  direct bearing on the relevant issues, see
  \citeN{stein-furth-consid-newt-meth,stein-meta-meth-newton}.}  That
framework allowed for the articulation of many different kinds of
theories, both wave and particle, and facilitated the switch from the
dominant particle theory to a new wave theory at the beginning of the
19th Century, at the hands of Young and Fresnel.  This framework,
however, is abductive in a subtler way than the ``standard form'' I
have sketched, at least in its actual historical applications.  There
was no biconditional between the concrete models and either particle
or wave theories up until, arguably, the 19th Century, when
diffraction decisively favored wave theories.  Even after the work of
Young and Fresnel, when the framework was pared down to accommodate
only wave theories, the available data did not allow the abduction of
a single, fully determinate theory until Maxwell's electromagnetic
theory of light was confirmed by Hertz's experiments on
electromagnetic radiation.  The derivation by \citeN{maxwell-dyn-emf}
of his full electromagnetic theory, abductive to its core, required a
framework far more comprehensive than Newton's for light and color,
though it did incorporate Newton's framework as a sub-framework, so to
speak.\footnote{Due to limitations of space, I cannot give a detailed
  argument that Maxwell's reasoning for his equations governing the
  electromagnetic field is abductive.  I will note here only that, in
  \citeN{maxwell-farad-line}, his derivation of the equations,
  including the necessity of the novel term representing the so-called
  displacement current, at bottom takes the form of a biconditional
  between, on the one hand, the electromagnetic phenomena observed and
  regimented by, \emph{inter alia}, {\OE}rsted, Amp\`ere, and (most of
  all) Faraday, and, on the other hand, what we now call Maxwell's
  equations, where the biconditional is implied by a Newtonian
  framework comprising the theory of a particular kind of Newtonian
  fluid.  His final, complete derivation in \citeN{maxwell-dyn-emf}
  also has this logical form, though the biconditional's antecedent is
  a Newtonian framework comprising a completely abstract theoretical
  representation of a medium whose dynamical behavior is governed by a
  Newtonian form of elasticity.  In \citeN{maxwell-treat-em}, he
  abductively derives the equations in the framework of Lagrangian
  mechanics.  I emphasize that these claims are crude and naive in the
  extreme, requiring detailed historical and technical exegesis for
  their complete elucidation and defense.  Nonetheless, I also claim
  that they capture the heart of the matter.  In the same vein, it is
  worth considering the argument of
  \citeN{hertz-bezich-maxwell-gegner} concerning Helmholtz's
  formulation of electromagnetism that he intended to be neutral
  between Maxwell's theory and Weber's action-at-a-distance theory (in
  other words, Helmholtz's formulation of a framework subsuming both).
  Hertz argued, in effect abductively, that Helmholtz's framework is
  in fact inconsistent with Maxwell's theory, in so far as it cannot
  represent the existence of free electromagnetic radiation, which
  itself abductively favors Maxwell's theory.  Again, a detailed
  defense of this claim is beyond the scope of this paper.}  Newton's
framework for light and color, therefore, in itself is perhaps best
described in its historical applications as supporting not the
abduction of theory from framework, but rather the abduction of
pared-down frameworks from a more general framework.  This all raises
fascinating and deep questions about the possible relations among
different frameworks, questions that I think are interestingly
different from similar ones about inter-theory relations, but which
are in any event beyond the scope of this paper.

I may have given the impression so far that abduction is a form of
reasoning peculiar to Newtonian mechanics (on the assumption that
Maxwell's abduction of his theory of electromagnetism was formulated
in a Newtonian framework, as I claim).  To the contrary, the history
of physics is replete with examples of abduction, and it is still used
today far and wide in all branches of physics.  Indeed, I do not think
it is an exaggeration to say that the large majority of revolutionary
advances in physics were based on abductive reasoning.  I do not have
the space here to make that case fully, so I will content myself with
listing exemplary instances across many branches of physics.
\begin{enumerate}
    \item In Lagrangian mechanics, one abducts the generic equations
  of motion by applying a variational principle (encoding the
  Euler-Lagrange equation, the abstract equations of motion) over the
  family of known individual models on the generic space of states.
  (I discuss this example in more detail in \S\ref{sec:fw-conf}.)
    \item In Hamiltonian mechanics, one derives the form of the
  Hamiltonian from the abstract form of Hamilton's equation in
  conjunction with the form of the phase portraits of the systems'
  evolutions in phase space (the individual models).  A trivial
  example: construct the phase portrait of a simple harmonic
  oscillator by continually measuring its position and momentum; the
  resultant set of points forms a circle in phase space (ignoring
  constant coefficients); it follows then directly from the abstract
  form of Hamilton's equation that the Hamiltonian must be
  $p^2 + q^2$, and by plugging this into the abstract form of
  Hamilton's equation one has abducted the generic equations of
  motion.
    \item In quantum mechanics, one abducts the form of the
  Hamiltonian from the abstract Schr\"odinger equation in conjunction
  with the paths of unitary evolution on Hilbert space (the individual
  models); the procedure is almost identical to that in Hamiltonian
  mechanics.  This is how the Hamiltonian for the Hydrogen atom,
  \emph{e}.\emph{g}., is derived.
    \item In quantum field theory, one abducts the form of the
  Lagrangian from observed symmetries, playing in this case the role
  of the concrete models \cite{sakurai-invar-princs-elem-parts}.
    \item In Boltzmannian statistical mechanics, one abducts the form
  of the Maxwell-Boltzmann distribution from the abstract equations of
  motion for statistical quantities, in conjunction with constraints
  imposed by empirically observed properties of the phenomena of
  thermodynamical equilibrium (concrete models)
  \cite{sommerfeld-thermo}.  The same holds for derivations of the
  microcanonical, canonical and grand canonical distributions in
  Gibbsian and quantum statistical mechanics
  \cite{fowler-guggenheim-stat-thermo-2nd,landau-lifschitz-stat-phys-1}.
  (\citeNP{malament-zabell-gibbs-phase-avg} provide a
  philosophically compelling explanation of the way that the physical
  characteristics of Gibbisan equilibrium for isolated systems are
  necessary and sufficient for the microcanonical distribution for
  appropriate physical systems; \citeNP{sorkin-mean-canon-ensemb}
  provides great physical insight into the way that appropriate
  further characterizations of equilibrium provide necessary and
  sufficient conditions for the canonical and grand canonical
  distributions respectively.)
    \item In ordinary thermodynamics, one abducts the principle of
  entropy non-decrease from the Kelvin Postulate (the abstract
  ``equation of motion'') in conjunction with the existence of
  irreversible processes, \emph{viz}., the concrete models
  \cite{fermi-thermo}.  This example shows that ``equation
  of motion'' can be broadly construed in the context of abduction, as
  I remarked in footnote~\ref{fn:eoms}.
    \item In general relativity, one abducts the FLRW cosmological
  models and the Schwarzschild solution from the Einstein Field
  Equation and the observed (or postulated) symmetries of the
  dynamical evolutions of those genera of systems
  \cite{wald-gr}.\footnote{It is a little delicate to explain the way
    in which a family of spacetime models such as the FLRW or
    Schwarzschild ones is relevantly like a set of generic equations
    of motion, but one can do this, and when one reconstructs the
    reasoning involved, say, in proving Birkhoff's Theorem (which
    implies the uniqueness of Schwarzschild spacetime given the
    assumed symmetries), it is indeed abductive in form.  If you
    object to the claim that general relativity is a framework, then
    think of this as the abduction of specific equations of motion
    from concrete models and generic equations of motion (general
    relativity considered as a theory, not a framework).}
    \item In cosmology, the inferences to dark energy and to dark
  matter, and so the construction of the entire standard $\Lambda$CDM
  (``cosmological constant plus cold dark matter'') model of
  contemporary cosmology, are abductive, starting from the framework
  of general relativity in conjunction with the concrete models
  constructed from observations such as the velocity-dispersion
  relations of galaxies, the isotropy of the cosmic microwave
  background radiation, and the large-scale accelerating expansion of
  the universe \cite{weinberg-cosmo}.
\end{enumerate}

In all of these cases, the abductive form of the entailment also
supports reasoning based on subjunctive conditionals, as in the case
of Newtonian gravitational theory.  This is almost trivially easy to
see in the cases of Lagrangian mechanics, Hamiltonian mechanics,
quantum mechanics, and quantum field theory.  In thermodynamics, it is
easy to show based on the abduction that, if there were no
irreversible processes, then the principle of entropy non-decrease
would hold only trivially (\emph{i}.\emph{e}., entropy would never
increase).  For the FLRW cosmological models of general relativity,
deviations from perfect symmetry (anisotropies, inhomogeneities) in
the concrete models subjunctively yield different spacetime models,
\emph{e}.\emph{g}., Bianchi-type models
\cite[ch.~22]{griffiths-podolsky-exact-sts} or Szekeres models
\cite{szekeres-inhom-cosm-mods}, depending on the form of the
deviations from symmetry.  In the $\Lambda$CDM model of cosmology, the
percentages of total mass-energy constituted by dark energy and dark
matter are subjunctively related to the values observed for
velocity-dispersion, accelerating expansion, and other such physical
quantities measured and represented in the concrete models of the
universe \cite{weinberg-cosmo,frieman-et-dark-energy-acc-uni}.  The
only case in which it is not clear to me that this is so is
Boltzmannian statistical mechanics.  Distributions for deviations from
equilibrium can be derived, but I do not know whether the form of
reasoning in that case supports the picking out of unique
distributions for given deviations from equilibrium.  I suspect it
does not.  It would be of interest to figure this out.

Now that the idea has been explicated and discussed, I conclude the
section by giving the promised sketch of an explanation of the way
that Newtonian abduction is similar enough in form and intent to
Peirce's original conception of abduction to warrant using the same
name for it.  As spelled out in
\citeN{peirce-ded-ind-hypo,peirce-nat-mean,peirce-abd-ind}, the
basic idea of abduction (or `hypothesis' as he sometimes calls it) is
that, from the assumption of a general rule $F$ and a basic state of
affairs $C$ relevantly related to the rule, a proposition $T$ is
derived by abduction in his sense if it is such that the conjunction
of $F$ and $T$ logically entails $C$, or else that $T$, on the
assumption of $F$, makes it probable that or offers a possible
explanation of $C$.  (Again, Peirce places no requirement of
``goodness'' on $T$ for it to be produced by abduction.)  If, in my
terms, $F$ is the framework (general rule), $C$ the concrete model
(basic state of affairs), and $T$ the theory (proposition), then the
similarity in form and intent between Peircean abduction and Newtonian
abduction is clear: in Newtonian abduction, the assumption of $F$ and
$T$ logically entails $C$, at the same time as guaranteeing that the
conjunction $F$ and $C$ entails $T$.  One can also view Newtonian
abduction as being similar to Peirce's original proposal in so far as
one might think that the theory derived by Newtonian abduction gives
an explanation of the phenomena that yielded the concrete model, given
that the phenomena is treatable by the framework.  Newtonian
abduction, therefore, differs from Peircean abduction primarily in
being logically stronger: the derived proposition in the former is
logically entailed by the ``premises'', which is not in general the
case in the latter.


\section{Framework Confirmation}
\label{sec:fw-conf}

In order to characterize the kinds of confirmation I claim accrue to
frameworks from abduction, it will be useful to begin with a
discussion of some observations on the topic by James Clerk Maxwell.
He is the only scientist or philosopher I know, besides Newton, who
explicitly recognizes Newtonian abduction as a separate form of
reasoning in science and articulates its form.

\citeN[p.~309]{maxwell-eom-conn-sys-in65} explains the idea
clearly:
\begin{quote}
  [T]he importance of [the] equations [of motion] does not depend on
  their being useful in solving problems in dynamics[,
  \emph{i}.\emph{e}., in deducing predictions of future behavior when
  the forces acting on a system are known].  A higher function which
  they must discharge is that of presenting to the mind in the
  clearest and most general form the fundamental principles of
  dynamical reasoning.

  In forming dynamical theories of the physical sciences, it has been
  a too frequent practice to invent a particular dynamical hypothesis
  and then by means of the equations of motion to deduce certain
  results.  The agreement of results with real phenomena has been
  supposed to furnish a certain amount of evidence in favour of the
  hypothesis.

  The true method of physical reasoning is to begin with the phenomena
  and to deduce the forces from them by a direct application of the
  equations of motion.
\end{quote}
It may seem that Maxwell contradicts himself here.  If we already know
the equations of motion---as we must, if we are to ``directly apply''
them to deduce forces from phenomena---then surely we know already the
forces as well, since the equations of motion are formulated in terms
of the forces, and only thence one deduces the phenomena, \emph{\`a
  la} HD.  The context of the passage, however, makes clear that
Maxwell has a conception of ``equations of motion'' in mind different
than that involved in employing a representation of known forces to
deduce resultant motions using generic or specific equations of motion
in conjunction with initial data.  What he calls here the ``equations
of motion'' are what I call, in essence, the abstract equations of
motion of a framework, and ``the fundamental principles of dynamical
reasoning'' are, in effect, the abstract structures of the framework
itself, as encoded in the abstract equations of motion and other
mathematical relations and principles, for example, that mass is
additive in Newtonian mechanics.

Maxwell began the paper (more of a brief note than a paper, really,
only two pages in length) by adverting to the method Lagrange
introduced in the late 18th Century in his \emph{M\'ecanique
  Analytique}, the essence of what we now know as Lagrangian
Mechanics, and contrasting it to that involved in the deduction of
dynamical behavior from (in my parlance) generic or specific equations
of motion in conjunction with initial data.  Recall that the
Euler-Lagrange Equation is the heart of Lagrangian Mechanics
\cite{curiel-cm-lag-not-ham}.  This equation results from the demand
that the dynamically possible evolutions manifested by a physical
system optimize, according to the principles of the calculus of
variations, a certain integral function of the dynamic quantities, the
action.  Thus, the Euler-Lagrange equation is, in my sense, the
abstract equation of motion in the framework.  To know the individual
models, then, of a genus of physical system (derived, say, by
identification with concrete models) allows one to abduct the generic
equations of motion as a particular instantiation of the
Euler-Lagrange equation: the variational principle entails that the
individual models are a particular family of paths on the space of
states if and only if the action is of a particular form (and so the
generic equations of motion are of a particular form).\footnote{For a
  graphic illustration of the method, see
  \citeN[\S4]{curiel-cm-lag-not-ham}, in particular the discussion of
  how one can read the generalized forces off from the form of the
  second-order vector fields on the tangent bundle of the
  configuration space for a given genus of physical system, those
  second-order vector fields representing the allowed dynamical
  evolutions of the system, \emph{i}.\emph{e}., its individual
  models.}  Thus, this method of deriving the equations of motion
exemplifies Newtonian abduction: knowing the concrete models, one
identifies them with dynamically possible paths on the space of states
(the individual models), and thence one abducts the form of the forces
acting on the system (the generic equations of motion) by applying the
machinery of the abstract equation of motion to the paths.  Such a
derivation does not so much provide evidentiary support for the
resulting generic equations of motion as being the most correct among
a field of competitors or something of the sort.  It rather
demonstrates, \emph{eo ipso}, that \emph{the derived equations have
  the structure appropriate for modeling the concrete dynamical
  evolutions of the system, and nothing more}---because the structure
of the derived generic equations of motion is directly determined,
biconditionally, by the structure manifest in those dynamical
evolutions (the concrete models, identified with the individual
models).

Newtonian abduction, therefore, shows, or captures, or elucidates the
mathematical structure manifest in the phenomena.  That part of the
world treated by the abducted theory, as investigated by appropriate
experimental techniques, directly evinces the abstract structure the
framework exhibits and requires.\footnote{One can therefore understand
  Newtonian abduction as providing grounds for the position of the
  ``sophisticated instrumentalist'' as characterized by
  \citeN{stein-yes-but}.  (I thank Tom Pashby for pointing this out
  to me.)  This does not, however, imply that Newtonian abduction by
  itself militates in favor of instrumentalism, nor realism either for
  that matter.}  A good example of this kind of structure is given by
the space of states and the dynamically possible evolutions of an
abstract Newtonian system, one treated by a theory whose generic
equations of motion are an instantiation of Newton's Second Law
\cite{curiel-cm-lag-not-ham}.  The algebraic and differential
structure of the Second Law itself imposes the structure of a vector
space on the family of all vector fields on the abstract space of
states representing possible interactions (``imposed forces'').  The
family of vector fields representing dynamically possible evolutions
of the system itself then accrues the structure of an affine space
modeled on that vector space.  It then follows that the abstract space
of states is naturally isomorphic to the tangent bundle over the
system's configuration space (\emph{i}.\emph{e}., there is an
isomorphism distinguished by the physics, in this case by the
existence of a privileged dynamical vector field representing ``free
evolution'').  Any theory abducted in the Newtonian framework inherits
these structures.  In particular, when one compares different concrete
models of the same system, say, ones constructed at different times
when it is experiencing forces of different magnitudes, one will find
the appropriate affine and additive structures on that family of
concrete models (when the concrete models are appropriately
interpreted using the conceptual machinery of the
framework).\footnote{It is important to note that no concrete model on
  its own has any of these structures; only rich enough families of
  them exhibit the structures as relations among them.}  This is easy
to verify in, \emph{e}.\emph{g}., Newtonian gravitational theory.
\citeN[p.~639, emphases his]{stein-struct-know} makes the point in a
trenchant way, explaining in illuminating detail how Newtonian
gravitational theory is ``a theory of \emph{a mathematical structure
  discernible in the world of phenomena, of observations, of
  experience}.''

What Stein says of Newtonian gravitational theory is straightforwardly
translated so as to apply to the entire Newtonian framework itself
(and indeed to all frameworks)---they provide mathematical structures
discernible in the world of phenomena, of observations, and of
experience, in such a way as to allow us to exploit those structures
for the theoretical representation of physical systems so as to
support substantive scientific reasoning of all forms about them.  To
perform empirically successful scientific reasoning by the direct and
ineliminable application of a theoretical structure, however, is to
endow that structure with some measure of confirmatory support, by any
reasonable standard.

I call this type of confirmation \emph{structural
  confirmation}.\footnote{After I finished this manuscript, I
  discovered that \citeN[pp.~208--209]{kuipers-strucs-sci-heur-patts}
  uses `structural confirmation' to refer to confirmation derived from
  instantiation of another relation, that of partial entailment,
  \emph{i}.\emph{e}., the probabilistic degree to which $A \vee B$
  entails $A$.  The two should not be conflated.}  It accrues to a
framework when one demonstrates that the structures intrinsic to the
framework are appropriate and adequate for representing and reasoning
about the genera of physical systems the framework purports to treat.
It is \emph{appropriate} if the framework's structures allow one to
identify in the relevant sense individual models of a theory in the
framework with concrete models of physical systems in the genus
treated by the theory.  It is \emph{adequate} if one can use that
identification to engage in substantive, successful scientific
reasoning about those physical systems, and, moreover, one has good
reason to believe that such identifications can be carried out for a
much broader range of relevantly similar systems than the ones already
treated.  Thus, when I follow Stein in speaking of ``a mathematical
structure [of a framework] discernible in the world of phenomena,'' I
mean that the framework facilitates the successful identification of
individual models with concrete models in such a way as to make it
possible for its theories to be used in successful reasoning.  In
particular, I make no claims that structures in the individual models
and ``in the world itself'' are ``isomorphic'' or ``homomorphic'' or
``similar'' in any way above and beyond the fact that the individual
models are relevantly identifiable with the concrete models, in large
part because those structures in the individual models do not form
part of the concrete models.

Before continuing the main thread of the discussion, I want to pause
for a moment to make clear what I am claiming such an identification
of individual with concrete models consists of, and in particular why
I claim that it neither presupposes nor implies any substantive
isomorphism between the two.  There are many ways one can legitimately
conclude that an individual model in a theory can be identified with a
concrete model constructed from experimental results.  In my
discussion of the example of the simple harmonic oscillator in
Newtonian mechanics, on p.~\ref{pg:sho-conc-mod}, in my introduction
of the idea of a concrete model, one identifies the individual model
with the concrete model by the brute comparison of the values in each
that one has some reason to believe represent the same physical
quantities.  It is true that, in this case, one identifies a geometric
circle in the concrete model with one instantiating an individual
model in the theory, and so there is isomorphism in this superficial
sense---circle to circle---but there is isomorphism of nothing else.
In particular, the individual model has far more structure than a
continuously parametrized sequence of values of physical quantities
(which exhausts the content of the concrete model), structure accruing
to it from the theory that makes it an individual model of \emph{that}
theory.  In the case of the simple harmonic oscillator, for example,
the individual model has the further implicit structure of the flat
affine structure of Newtonian spacetime, which defines the derivative
operator with respect to which the individual model is a smooth curve.
The concrete model need not have that structure.  It really is just a
continuously parametrized sequence of values of physical quantities.
Even if one were to put a flat affine structure into the concrete
model by hand, it would be superfluous, as the identification of the
concrete model with the individual model does not depend on it.  To
take the isomorphism between the two continuously parametrized
sequences of values of physical quantities---the only isomorphism
available---as a philosophically substantive isomorphism would be
nothing more than the sheerest naive verificationism.

In consequence, one can agree with the idea of structural confirmation
while remaining agnostic about all issues pertaining to realism and
anti-realism, as any good confirmatory relation should allow.  In
particular, structural confirmation has no necessary connection to the
idea of structural realism (though I suspect that champions of the
program may want to try to avail themselves of it in their arguments).

It may sound strange to say that Lagrangian mechanics itself admits of
confirmation, but I claim it does.  Structural confirmation does not
give us a reason to believe that Lagrangian mechanics is true.  I am
not even sure what that claim could mean.  Structural confirmation
rather gives us reason to believe that the structures intrinsic to
Lagrangian mechanics are approriate and adequate for representing many
genera of physical systems.  We come to believe in its fruitfulness as
a ground for successful scientific reasoning of all sorts.  That,
however, is exactly what standard confirmation does for us with regard
to ordinary theories.  It is only the realist who believes
confirmation also gives a reason to think the theory is true; the
instrumentalist is not committed to taking that further
step.\footnote{Of course, everything said of structural confirmation
  of a framework holds as well, \emph{mutatis{} mutandis}, for a
  theory.  Theories also are amenable to structural confirmation, in
  the same way, by abductively showing that the structures intrinsic
  to the theory are appropriate and adequate for representing and
  reasoning about the concrete models of the different species of
  physical system the theory purports to treat.}

Abduction shows that a genus of physical systems is governed by the
generic equations of motion and other mathematical relations imposed
by a theory in the relevant framework if and only if the relevant
space of states and family of dynamically possible evolutions,
appropriately identified with the family of concrete models, has that
structure.  In other words, those systems are appropriately and
adequately treated by a theory in the framework if and only if their
predicted behavior, both in isolation and in response to the kinds of
interactions with the environment allowed by the structures of the
theory, accords with the constraints imposed by those structures.
This is not only a successful prediction of the theory by the
framework---it is also the provision of grounds for the possibility of
using the theory to make successful predictions, \emph{i}.\emph{e}., a
necessary precondition for being able to apply the theory in that way,
when the theory is generated by abduction.  As such, every successful
application of the theory in, \emph{e}.\emph{g}., making predictions
of the HD-type, endows confirmation on the framework as a whole, when
the theory is generated by abduction, since abduction made it possible
in the first place.

Newtonian abduction is a relation between theory and experiment that
standard accounts of confirmation based on HD-type reasoning cannot
accommodate.  Structural confirmation, nonetheless, as I just
intimated, derives in part from the validation of a proposition that
itself entails a proposition largely the same in form as used in the
standard accounts of confirmation based on HD forms of reasoning: $H$
(the framework) entails some $E$ (evidence, in this case the
biconditional between theory and concrete models).

Indeed, with the addition of two extra assumptions---determinate
values for the constant coefficients in the generic equations of
motion to turn them into specific equations of motion, and initial
data---, and replacing the concrete model in the abductive proposition
with the individual model derived from the specific equations of
motion and the initial data, the abductive proposition even implies a
proposition identical to that used in HD, by discharging one of the
directions of the biconditional: the framework entails that $H$ (an
abducted theory) entails $E$ (evidence in the form of an individual
model that can potentially be identified with a concrete model).
Nonetheless, as the description of the way that one can use the
abductive proposition to derive one as used in HD makes clear,
concrete models play no role in the logic of HD itself, as I remarked
in \S\ref{sec:abd} just after formula~\eqref{eq:log-form-hd}.  HD
predicts an individual model.  Concrete models come into play only
after the logical derivation of the individual model, in the attempt
to identify the deduced individual model with a concrete model, to
test the HD prediction.  This is why Newtonian abduction automatically
shows the theory to be appropriate and adequate for treating the
phenomena that yielded the concrete model in a way that HD can never
do, since the concrete model itself forms an essential part of the
reasoning.  Since the identification of individual model with concrete
model occurs in HD only after the reasoning has concluded (if it is
possible at all), there can be no such necessary conclusion about the
propriety of the theory in treating the phenomena.

I suspect that one may be able to extend some of the standard accounts
of confirmation, based on HD reasoning, to accommodate Newtonian
abduction and structural confirmation, but the conceptual obstacles
are formidable.  There are as well obvious and \emph{prima facie}
difficult technical obstacles that immediately present themselves.
For instance, in any account of confirmation that quantifies it using
probabilities, such as Bayesianism, one will have to construct a
probability space that includes as elements the framework itself, the
theories possibly abductable from the framework, the individual models
of those theories, and the concrete models that may be appropriate for
identification with those individual models.  That is not all,
however.  As the discussion of Newton's and Le Verrier's use of the
Precession Theorem shows, the probability assignments on the space
will have to respect the non-trivial relations \emph{among} the
theories (topological, algebraic, differential, and so on) that the
framework endows on its family of theories, those relations that make
Newtonian abduction possible in the first place and are integral in
its support of reasoning based on subjunctive conditionals (and so
integral to what I call \emph{modal confirmation}, discussed below).
These relations go far beyond the standard $\sigma$-algebra required
on a space in which one defines a probability measure satisfying the
Kolmogorov axioms.\footnote{Indeed, structural \emph{evidence} of the
  kind relevant to abduction in general differs in important ways from
  ``raven counting'' evidence and from HD predictive evidence,
  primarily because of the non-trivial mathematical relations among
  theories one must account for in formulating and analyzing the
  evidence.  A general discussion is beyond the scope of this paper.}
One will have to show, moreover, that the further structures are
consistent in the right ways with the probability structure.  These
are highly non-trivial problems.\footnote{\label{fn:technical}To get a
  sense of the difficulty and depth of the technical problems one
  faces, consider the family of all possible ``theories'' one can
  construct in the Newtonian framework, \emph{i}.\emph{e}., all (say)
  twice-differentiable functions (force laws) on Newtonian spacetime.
  This can not even be turned into a topological manifold, since the
  different force laws can depend on any number of different variables
  (position, velocity, fluid flow, shear-stress, and so on), and so
  are functions with infinitely many different possible domains.
  Restrict attention, therefore, to the family of all possible
  theories for a fixed set of physical quantities (say, those
  appearing in Navier-Stokes theory).  This, then, forms an infinite
  dimensional vector space, modeled on the field of real numbers,
  which can be turned into a Banach space by choice of a norm; there
  is, however, no obvious, unique, physically significant choice of
  norm, though many are available (\emph{e}.\emph{g}., the
  $\sup$-norm, and so on).  There is as well no obvious, unique,
  physically significant topology to put on the space, though many
  topologies are available (compact-open, Whitney, one based on the
  $\sup$-norm, and so on).  One now wants to put something like a
  probability measure on it that respects the topology, in the sense
  that small perturbations don't drastically change the measure of
  open sets, so one wants a Borel measure that is, if not invariant
  under linear translations, then uniformly bounded in some way by the
  size of the translation.  There is, strictly speaking, no
  probability measure on such an infinite-dimensional space, and no
  translation-invariant Borel measure on it at all
  \cite{curiel-meas-topo-prob-cosmo}.  It is an open---and
  difficult---question whether one can construct such a measure that
  is uniformly bounded under translations.  Even if one were to
  construct such a measure, it would necessarily assign infinite size
  to the space as a whole, and so one would need some sort of
  regularization scheme to extract meaningful probabilities for
  non-trivial open sets.  See \citeN{curiel-meas-topo-prob-cosmo} for
  a discussion of how to try to do this in the context of defining
  physically significant probability measures in cosmology, and the
  many further problems that arise.

  One may be able to get the technical machinery under control if one
  can ``localize'' the family of laws to an appropriately restricted
  subfamily.  Consider the case of Newton's and Le Verrier's use of
  the Precession Theorem to support the inverse-square form of
  Newton's gravitational law.  The ``state space'' of laws now
  consists of functions of the form
  $\displaystyle \frac{1} {r^{2 + \epsilon}}$, for small
  $\epsilon \ge 0$ and, say, $\le .5$.  This naturally forms a compact
  1-dimensional manifold with boundary, with a natural, physically
  significant metric function on it (the absolute difference in
  $\epsilon$).  There is thus an obvious, physically significant
  topology to put on the space, based on the metric.  The absolute
  difference in $\epsilon$ gives the space a locally affine structure:
  if $\displaystyle \frac{1} {r^{2 + \epsilon_1}}$ and
  $\displaystyle \frac{1} {r^{2 + \epsilon_2}}$ are in the space
  ($\epsilon_1 < \epsilon_2$), then so is
  $\frac{1} {r^{2 + \epsilon'}}$ for all
  $\epsilon' = \lambda \epsilon_1 + (1 - \lambda) \epsilon_2$, for
  $\lambda \in [0, \, 1]$.  One now wants to put a probability measure
  on it that respects the topology and the affine structure, in the
  sense that small perturbations don't drastically change the measure
  of open sets.  One therefore wants a Borel measure that is invariant
  under affine translations, or at least uniformly bounded in some way
  by the size of the translation.  There are many such measures.}

In sum, abduction is a form of reasoning that indubitably shows that a
theoretical apparatus can be applied to empirical data in such a way
as to produce, in the context of the apparatus, representations of the
relevant systems that are of necessity predictively accurate; those
representations, moreover, are automatically fruitful in the sense
that they are instantly ready to apply to the attempted representation
of further systems one has reason to believe are of the same genus.
The systematic production of successful theoretical representations of
the nature and behavior of physical systems, however, is the most
fundamental and most important form of confirmation, whatever form
that production may take.

Structural confirmation is a far stronger and more substantial form of
confirmation than that provided by merely deducing predictions and
matching them against observations, as Maxwell himself emphasized.  As
he says, ``A higher function which [the framework] must discharge is
that of presenting to the mind in the clearest and most general form
the fundamental principles of dynamical reasoning.''  This is what
gets confirmed---not just the abstract structure of the framework
manifest in the phenomena, but all the fundamental principles of
dynamical reasoning that are part of the framework and allow one to
\emph{apply} the framework to the experimentally characterized
phenomena.  Structural confirmation, therefore, provides us more than
just good reason to accept the framework as a useful theoretical tool
for the scientific investigation of parts of the physical world (or,
in more standard parlance, does more than give us good reason to have
credence in the framework).  Structural confirmation substantiates the
fruitfulness of the framework as a system within which one has good
reason to believe that one can formulate theories that automatically
inherit a high degree of confirmatory support in the standard sense.

The fact that Newtonian abduction directly supports the formulation of
and arguments based on subjunctive conditionals provides further
confirmatory support of a sort that standard accounts based on HD
cannot.  It is, again, an instance of the framework's making possible
more elaborate and convincing argumentation in favor of a theory's
empirical success than HD by itself, or any other form of scientific
reasoning standardly treated in accounts of confirmation, makes
possible.  The production and use of subjunctive conditionals,
exemplified by Newton's and Le Verrier's use of the Precession
Theorem, is confirmatory of the framework itself, because the
derivation of the relevant subjunctive conditionals is possible only
by use of the framework itself as whole, not any individual theory in
isolation, since the conditionals are comparative among different
theories.  It is exactly the topological, algebraic, geometric,
differential, and other structures that the framework imposes on its
family of theories that permits this kind of reasoning.  Reasoning of
the HD type cannot accomodate this, because it deals with theories in
isolation from each other.  It has no place for reasoning based on
structures accruing to a family of theories.\footnote{I do not claim
  that standard forms of HD reasoning, and standard accounts of
  confirmation based on them, cannot be modified or extended to cover
  the form of reasoning I discuss here.  I claim only that no account
  I know of in its present form has the capacity to do so.  It would
  be an interesting project to attempt to modify or extend extant
  accounts to try to do so.}

For the same reasons, confirmation accrues to the framework itself by
the further use of such subjunctive reasoning to construct sequences
of ever more detailed and accurate individual models, as Newton and Le
Verrier did in calculating the orbital perturbations due to
inter-planetary gravitational effects, and similarly for Newton's
treatment of comets, showing that they all accorded with the
predictions of the theory even more accurately than the initial
individual models did
\cite{harper-newtons-sci-meth,smith-closing-loop}.  That the
framework supports such successful reasoning, employing the theory
itself in powerful ways to directly yield more accurate
representations based on abductive arguments, confirms the empirical
propriety and adequacy of the framework.  I call confirmation based on
the formulation and successful use of subjunctive conditionals
\emph{modal confirmation}.  I will use \emph{confirmation by Newtonian
  abduction} to refer to both structural and modal confirmation, when
what I say applies to both.

Similarly, when the predictions of a theory fail and modifications of
its individual models cannot mitigate the failure, the framework will
provide structures, concepts and methods for guiding us in trying to
figure out why it failed, and possibly in trying to formulate new
theories by abduction that will provide accurate representations of
the concrete models.  Newton's framework for theories of light and
color, for example, did this in grounding the move from a particle to
a wave theory of light at the beginning of the 19th Century.  If a
framework does so successfully, this also provides confirmation, for
it is based on its abductive character: the framework can, in the best
cases, tell one what other theories or sorts of theories to look for
and examine, and when the search is successful, then confirmation by
Newtonian abduction accrues to the framework.  This is
methodologically richer than HD, and each such success compounds the
confirmation accruing to the framework.

Structural confirmation is in itself non-contrastive.  Confirmation
does not accrue to the framework by being stronger or more appropriate
than rivals, and one does not need to worry about ``unconceived
alternatives'' \cite{stanford-unconc-alters}: one has shown that the
framework does everything one could possibly ask of it; asking more
would be to demand the supererogatory.  Of course, one is still free
to compare the abductive power of different frameworks, when more than
one is available in a context in which theories can be abducted
possessing individual models appropriate for identification with the
same family of concrete models.  There are then (at least) three ways
that one of the frameworks can garner more confirmation from abductive
success than the other.  First, one of the frameworks may contain a
theory whose individual models apply to the given family of concrete
models and to others in addition, which the other framework's theories
cannot.  Second, the individual models of the theories of one of the
frameworks may be predictively more accurate than those of the other.
Third, the individual models of one of the frameworks may ground the
possibility for novel predictions that the other does not, and those
predictions can then be confirmed.  All three conditions obtain for
general relativity as compared to Newtonian mechanics (in so far as we
are able to test separately the predictive accuracy of theories
formulated in each framework against the same concrete models, which
we cannot in fact always do).\footnote{I am glossing over a subtlety
  here.  The concrete models one compares to the individual models of
  each framework will not always be the same, but will rather
  sometimes have to be constructed from the raw data in different ways
  using the different theoretical concepts and structures of the
  different frameworks, so as to fit into the different abductive
  propositions of each framework.  That is not in general a problem.
  In such cases, the individual models of one can be ``translated''
  into those of the other in a way that preserves enough physical
  significance for one to have confidence that the individual models
  in the different theories represent the same dynamical behaviors of
  the same physical systems.  Such translations often take the form of
  limiting or approximative constructions.  For example,
  \citeN{malament-newt-grav-geom-spc}, in effect, shows how to do this
  for general relativity and Newtonian gravitational theory.  If one
  cannot construct such translations, then one has no reason to
  believe that the two theories are representing the same behaviors of
  the same physical systems.  Nonetheless, it is sometimes the case
  that the same concrete models can in fact be identified with
  individual models of different frameworks, as in the case of
  Keplerized planetary orbits and individual models in general
  relativity (Schwarzschild spacetimes with test particles traversing
  geodesics in a 3+1 representation of spacetime) and Newtonian
  mechanics (two-body solutions of Newtonian gravitational theory).}
It also follows that, by any measure, a higher degree of confirmation
should accrue to a framework that generates more theories that are
successful than to one that generates fewer over the same domain of
physical systems.

I am not claiming that a physical theory has strong confirmation only
in so far as one can cast it in Lagrangian form, or only in so far as
one has an abductive derivation of its equations of motion from a
collection of concrete models constructed from experimental or
observational results.  I quoted Maxwell on the Lagrangian framework
and discussed it at length only to drive the point home, as it comes
out with such marvelous clarity there: a theory is successful in
representing a genus of physical system in the epistemically strongest
way, with the highest degree and quality of confirmation, when the
structure of the theory's generic equations of motion not only
conforms to but embodies the structure manifest in the phenomena it
purports to represent, that is, when the structure of the theory
recapitulates the structure manifest in the physical evolutions of the
system and as little else as possible.  That is what the biconditional
at the heart of Newtonian abduction guarantees.\footnote{I believe
  that Newton's infamous proclamation, ``Hypotheses non fingo,'' is
  best understood in the context of his characteristically abductive
  form of reasoning: whatever is not derived abductively from the
  phenomena is a hypothesis, and those he avoids as much as possible,
  precisely because, being merely postulated and then tested by weaker
  forms of reasoning (such as HD or induction), they cannot accrue the
  same degree of certainty as propositions derived by abduction.}

One may be concerned that my arguments seem to require that
methodological and epistemological principles such as ``rules of
evidential warrant'', ``standards of good argumentation'', and so on,
be such as to admit confirmation, which it seems they must if they are
to be included in the framework and the framework as a whole admits
confirmation.  I claim that it does make sense to treat them as
susceptible of confirmation, in the sense that one has reason to use
and trust them after they have been shown to play an ineliminable role
in successful scientific reasoning.  (I say more about this in
\S\ref{sec:true-confirm} below.)  One must be careful, however, in
one's judgement of what ``belongs to a framework'' and what does not.
When I speak, \emph{e}.\emph{g}., of ``principles of good
argumentation'' as forming part of a framework, I do not mean to imply
that I think that first-order predicate logic thereby must be part of
it (and is thereby susceptible to confirmation).  I mean rather: forms
of argument \emph{peculiar to the domain of physical systems the
  framework purportedly treats}.  For some argumentative purposes with
regard to a given type of physical system, some ``standard'' forms of
mathematical and heuristic argumentation will not be appropriate, even
though they work well in other fields of physics.

An example is how perturbations are treated, when understood as
leading terms in a (truncated) power expansion in some dynamically
relevant parameter.  That works well for systems that are well behaved
in an appropriate sense, but not for others, \emph{e}.\emph{g}.,
chaotic systems.  Thus, the argument in the 19th Century that the
Solar System is stable because it is so up to second order in a
standard expansion representing the perturbative effects of the
planets' gravitational forces on each other, though unobjectionable in
and of itself as a completely standard form of perturbative reasoning,
actually fails, because, as Poincar\'e showed, the third-order terms
exhibit chaotic instabilities.  The lesson is that one must be
extremely careful in applying perturbative analysis to any Newtonian
gravitational system of more than two bodies.  The recognition and
respecting of the required care is part of the canon of ``good
argumentative forms'' for that theory.  Another good example is the
use of the linearized Einstein field equation as a truncated
perturbative expansion.  One must be careful in ascertaining that the
phenomena at issue are in the appropriate ``weak field regime'' before
one one is justified in using the linearized form in general
relativity.  Similarly, with regard to standards of what can count as
good evidence, only what is peculiar to the framework or theory forms
a part of it in my sense.  An example is stellar aberration.  Before
telecopic observational prowess, in conjunction with calculational
capacity, had developed to such an extent that the phenomenon could be
measured and controlled for, worrying about it did not form a part of
good evidentiary standards peculiar to astronomy; afterward, it did.

Because it is the only work in the literature I am aware of that more
or less directly addresses the issues I investigate here, before
concluding this section I will discuss
\citeN{henderson-et-sci-thrs-hier-bayes}.\footnote{\label{fn:ibe}Works
  such as \citeN[ch.~7]{lipton-ibe} and \citeN{henderson-bayes-ibe}
  are not relevant to my arguments, as their discussions of abduction
  treat it as IBE.  Lipton does talk about the capacity for IBE{} to
  confirm entire frameworks (\emph{e}.\emph{g}., on p.~60, in his
  discussion of Newtonian mechanics and special relativity).  His
  reasons for saying this, however, are based on
  theoretical---non-empirical---virtues, such as unification, scope
  and explanatory power, that he claims accrue to frameworks based on
  their role in IBE.  As should be clear, my arguments that Newtonian
  abduction can confirm frameworks derive directly from the logical
  form of the reasoning itself in its immediate contact with
  \emph{empirical} evidence.}  I think their analysis is interesting
and potentially fruitful in many ways, but I will here focus on its
shortcomings with regard to the issues I am concerned with.

Their account distinguishes between higher and lower level theoretical
systems, which they at different times and in different contexts refer
to as `frameworks', `theories', and `hypotheses', for both lower and
upper level systems.  (Nothing hinges on the differences in
terminology.)  I shall use `framework' for a higher-level theoretical
system in their account and `theory' for a lower-level one, which may
be generated by a framework.  One of the virtues of their account is
that they are usefully ambiguous about what ``generation'' of a theory
from a framework may be---it may be deductive entailment, or
constructive approximation, or it may instantiate some other relation
entirely.  Thus nothing \emph{prima facie} rules out that their
analysis may be relevant to Newtonian abduction.

In the account, each level is conceived of as an unstructured set of
mutually exclusive alternatives.  This, however, from the start misses
out on several important features of the relationships among competing
frameworks and those among competing theories in a single framework.
A framework in general endows its associated family of theories with
rich mathematical structures (topological, geometric, algebraic, those
grounded in real or complex analysis, and so on), all with manifest
physical significance, as was seen in the discussions of the
Precession Theorem and modal confirmation.  Nothing prevents
Henderson, \emph{et al}., from endowing each level of a hierarchy with
such structures, but neither would anything in their account allow one
to exploit those structures, because of the strong sense in which they
conceive of theories as being mutually exclusive: the relationship of
one theory to another, in their setup, cannot tell us anything about
the confirmatory support of either theory.  This restriction entails
that their account cannot represent and account for in an accurate way
some of the strongest possible confirmatory evidence for a framework
or a theory, \emph{viz}., modal confirmation arising from the
successful use of subjunctive conditionals based on abductive
reasoning.  I see no way around this problem in their approach.  It is
important to note, however, that this problem is not peculiar to their
account.  It infects any account of confirmation in which theories are
represented as independent in this strong sense, such as in the
Bayesian one (which Henderson, \emph{et al}., work in).

In any event, it is difficult to see how abductive confirmation of the
sort I explicate here can be represented at all in their account.  Let
$T$ be a lower-level theory, $F$ a higher-level one (a framework) that
generates it, and $D$ some structured data that stands in direct
evidential relation to $T$.  They consider the probabilities
$P(T | F)$, and $P(F | D)$, but they never consider $P(F | T)$ or
$P(F | T \wedge D)$, much less $P(F | T \equiv D)$, which represent
the fundamental relations I am concerned with.  In fact, it is not
even clear to me what $P(T | F)$ \emph{means} in their account, since
the two theoretical systems live in different probability spaces
(since each level of the hierarchy has its own probability space).  In
so far as one can make sense of $P(T | F)$ in their approach, however,
then nothing further in their discussion \emph{prima facie} seems to
preclude a discussion of $P(F | T)$, $P(F | T \wedge D)$, and
$P(F | T \equiv D)$ in the same vein as theirs, but neither is it
clear to me how such an analysis would go, nor how one would justify
it in the terms of their arguments, especially in light of the
difficulties I pointed out in the previous paragraph (and those I
discuss in footnote~\ref{fn:technical}).  It would be interesting to
investigate whether or not this can be done.

Finally, on their account higher-level frameworks do not gain direct
confirmatory support from data, but gain it only indirectly, mediated
by the theories they generate that are confirmed by the data (p.~185).
This misses out on the fundamental logical structure of Newtonian
abduction that lends direct and immediate confirmation to the
framework itself from the empirical evidence: the data and the theory
enter into the confirmatory relation in a symmetrical, equivalent way
(the biconditional component of the abductive proposition), and so the
structured data itself in its instantiation of the framework's
abstract structure lends the framework confirmatory support as
directly as does the success of the theory in producing individual
models identifiable with the concrete models in a predictively
accurate way \emph{\`a la} HD-confirmation.

Nonetheless, I emphasize again that I do not think that
\citeN{henderson-et-sci-thrs-hier-bayes} has nothing in its
favor.  To the contrary, I find the approach compelling in many ways
and a potentially rich source of ideas and machinery for approaching
many important problems in confirmation theory.  It is also useful for
my particular purposes here: in so far as their analysis is
successful, it shows another way that confirmation can accrue to
frameworks besides by Newtonian abduction.  I wanted only to
underscore the differences in aim, in detail of implementation, and in
consequence between their work and the present project.

\citeN{romeijn-abd-bayes} is another interesting case.
Although the subject of his arguments is the construction of a
Bayesian model of abduction as, in effect, IBE, his analysis is
general enough to cover all forms of the generation of
``suppositions'' from data, and so arguably encompasses at least part
of the meat of Newtonian abduction.  He says (p.~437),
\begin{quote}
  What the above examples show is that if we happen to choose the
  theoretical concepts well [by IBE], then they will be beneficial to
  the predictions, instead of being irrelevant or detrimental.  In
  other words, the examples show that theoretical notions have an
  evaluative rather than a generative use.
\end{quote}
This harmonizes well, in obvious ways, with my discussion of Newtonian
abduction, so far as it goes.  Newtonian abduction, however, allows
one to go further than Romeijn's insightful analysis does: the logical
entailment of the biconditional between the theory and the concrete
models---this biconditional being the analogue of Romeijn's ``choice
of theoretical concepts'', when the framework is already in
place---supports confirmation of the framework itself.  Romeijn is, I
think rightly, reluctant to conclude that his own analysis supports
that conclusion.

His analysis also illuminates how Newtonian abduction is superior to
IBE in one important way.  As he observes (p.~437),
\begin{quote}
  [N]othing in [my arguments] answers to the problem [with IBE] that
  there are infinitely many theoretical notions that are potentially
  useful [in attempting to model given phenomena], and that we do not
  have any reason to choose any particular one when learning from the
  observations.
\end{quote}
Within the fixed context of a framework, when Newtonian abductive
reasoning is available, this is false: if a framework abducts a
successful theory from the phenomena, then one has the strongest
possible grounds for choosing the framework, and choice of the
framework just is Romeijn's ``choice of theoretical concepts''.  When
``choice of theoretical concepts'' means choice of theory in a
framework rather than choice of a framework itself, Romeijn's
pessimism is again unfounded when Newtonian abduction is available,
since that picks out the unique theory in the framework best suited to
represent the phenomena.  When there is more than one framework
available for the representation of a given family of concrete models,
moreover, and both support the abductive generation of a theory to
represent those models, then one may have available the kind of
comparative confirmation of the frameworks I discussed above, on the
basis of which to choose in a principled way between them.  This, by
the way, also rebuts the ``argument from a bad lot'' proposed by
\citeN{fraassen-laws-sym}, at least in the context of Newtonian
abduction, again showing its superiority to standard IBE.

As I mentioned in footnote~\ref{fn:theory-med-meas}, agreeing
determination of the values of parameters by theory-mediated
measurements, as characterized by \citeN{harper-newtons-sci-meth} and
\citeN{smith-closing-loop}, is not an HD prediction but rather
has the form of a Newtonian abduction.  In these cases, different
theories are used to determine the value of some physical parameter,
and the agreement in values derived from different theories provides
strong evidence in favor of that value, at the same time as it
provides confirmatory support for the theories themselves.  An
edifying example is the determination of the value of the cosmological
constant $\Lambda$ in the standard $\Lambda$CDM model of contemporary
cosmology, by way of concrete models constructed from observations of:
the distribution of the velocities of supernov{\ae}; the spectrum of
primordial mass-energy density amplitude-fluctuations encoded in the
cosmic microwave background radiation (CMBR); the distribution of
anisotropies in the CMBR; the ratio of the mass of x-ray emitting gas
in galaxies to total galactic mass; and the observed age of the oldest
stars compared to the observed age of the beginning of large-scale
structure formation.  (See
\citeNP{frieman-et-dark-energy-acc-uni}.)  Each of these
observations yields a value for the cosmological constant by reasoning
of the form of a Newtonian abduction, each in a different theory.  The
fact that all such determinations agree manifestly provides
substantial confirmation both for the value of the parameter and for
the theories themselves.  The only structure all these different
theories have in common, however, is the framework of general
relativity in which all the theories are formulated, by which they are
each able to make substantive contact with the $\Lambda$CDM model in
the first place to derive the value of $\Lambda$ from their respective
concrete models, and by virtue of which one can identify the $\Lambda$
in each theory as representing the same physical quantity as those in
the others.  In so far as confirmation accrues to the individual
theories based on their role in the reasoning that determines the
value of the parameter, so must confirmation accrue to the framework
as well, for without the framework one could not even compare the
different values in a meaningful way: the shared framework guarantees
that the different theories are representing the \emph{same} physical
quantity, and so the framework itself plays an ineliminable role in
the abductive reasoning resulting in the agreeing measurements.

Finally, as I noted in footnote~\ref{fn:condl-defn}, Newtonian
abduction has the form of a Carnapian conditional definition
\cite[pp.~441\emph{ff}.\@]{carnap-test-mean-1}, a form of reduction
sentence.  Carnap, and later philosophers such as
\citeN[pp.~23\emph{ff}.\@]{hempel-funds-conc-form}, considered the
uses of this logical form only with regard to the definition of
theoretical terms in a formal scientific language by reduction in a
technical sense to observational terms in that language, not, as here,
as the basis for a distinctive form of scientific reasoning.  Still,
the comparison is illuminating.
\citeN[p.~26]{hempel-funds-conc-form}, for example, says,
\begin{quote}
  A reduction sentence [of the conditional form] offers no complete
  definition for the term it introduces, but only a partial, or
  conditional, determination of its meaning; it assigns meaning to the
  ``new'' term only for its application to objects which satisfy
  specific ``test conditions.''
\end{quote}
The ``test conditions'' Hempel refers to are the analogue of the
concrete models appearing in the biconditional of the abductive
proposition~\eqref{eq:log-form-nabd}.  Thus, in a similar vein, one
may construe the abductive proposition as the determination of the
partial or conditional validity of the theory, and thereby the
determination in part of its empirical content and meaning, when the
``test conditions'' constituted by the phenomena that yielded the
concrete model in conjunction with the framework are satisfied.  In so
far as the framework provides the theoretical basis for the
attribution of empirical content to a theory, and in so far as the
framework acts, so to speak, as the constitutive \emph{a priori}
component of that empirical content, confirmation of the theory by
validation of its empirical content should flow upwards to lend
confirmation to the framework as well.

\section{This Is True Confirmation}
\label{sec:true-confirm}

When I presented the material of this paper in a talk at the
conference ``Reasoning in Physics'', at the Center for Advanced
Studies at Ludwig-Maximilians-Universit\"at in Munich, in December
2016, a sizeable part of the audience objected that Newtonian
abduction does not provide real confirmation to frameworks, but rather
articulates only an explanatory or grounding relation among
frameworks, theories and structured data.  First, I contend that, in
so far as the abductive proposition implies a proposition having the
same logical structure as HD reasoning as used in standard accounts of
confirmation, and that the empirical content of the terms in that
proposition are essentially the same as in the standard HD case, the
presumption must be strong that Newtonian abduction provides
confirmation of frameworks at least in that same way, even if in no
other way.  Second, even if my analysis does articulate an explanatory
or grounding relationship, then \emph{eo ipso} the relationship is
confirmatory in the standard sense: to show that a theoretical system
explains or grounds the success of our theoretical representations of
the world by itself provides confirmatory support for the system in
the standard sense \cite{tesic-et-conf-explan-bayes-ibe}.

As I argued in \S\ref{sec:fw-conf}, however, the type of confirmation
that Newtonian abduction confers on frameworks goes beyond that of
standard accounts.  It provides confirmation by showing how an
abductively successful framework can give us \emph{understanding} of
the physical world---we come to see that and how the phenomena in the
world of our experience manifest the abstract structure of the
framework.  By ``understanding'' here, I mean something like ``the
capacity to operate successfully in the scientific enterprise, in all
its forms, aspects and parts''---to use our theoretical
representations and experimental practices and results as the basis
for the fruitful continuation of the enterprise: as part of evidential
warrant in testing; as basis for characterizations of systems and
predictions about them; as inspiration for potentially fruitful new
investigations; as the grounds for conceptual clarification and
innovation in foundational work; and perhaps most of all to grasp how
our theoretical representations and reasoning on the one hand and our
experimental practices and results on the other relate to, inform, and
substantively contribute to the constitution of each other, and to
grasp that in such a way as to ground their use \emph{in} the fruitful
continuation of the enterprise.  To come to understand the phenomena
in this way---to grasp that they manifest the structure of the
framework---\emph{eo ipso} gives one the capacity to do all this, by
guiding one's use of the framework in further investigation both of
the phenomena one has already investigated and of other systems one
has reason to believe are appropriately similar.  Deduction of the
individual models from equations of motion cannot do this, for it can
show at most that the phenomena manifest a structure compatible in
some way with that of the equations.  It is the entailment of the
biconditional between the theory and the concrete models that shows
the phenomena to manifest exactly that structure.\footnote{I want to
  emphasize again that none of this has anything to do with any issue
  pertaining to realism and anti-realism.  There is no claim made or
  needed that the structure manifest in the phenomena,
  \emph{i}.\emph{e}., the structured data, is ``really'' part of the
  furniture of the world, in some deep metaphysical sense.  That it is
  manifest in the phenomena, in the sense that one can identify the
  structured data the phenomena yields with individual models,
  suffices for the soundness of the evidential and confirmatory
  relations at issue.  Those relations are agnostic about realism, as,
  again, any good confirmatory relation should be.}

Newtonian abduction is superior in this way as well to IBE,
\emph{viz}., with regard to how it provides understanding and
explanation.  As \citeN[p.~437]{romeijn-abd-bayes} makes clear, it is
difficult to see how IBE supports any substantive understanding of why
a theory is predictively accurate for a class of phenomena, nor of why
the theory explains the phenomena.  Newtonian abduction, to the
contrary, shows how we may come to a deeper understanding of the
physical world even in the complete absence of explanations: Newton
developed the theory of universal gravity by explicitly
\emph{abstaining} from giving any ``explanation'' of the origin of the
force, mechanical or otherwise.  It would, nonetheless, be folly to
deny that the development of that theory did not contribute in a deep
and profound way to our understanding of the world.

Much of the empiricist tradition over the past 90 or so years in the
study of the relations between data and theory have led to skeptical
claims and questions about the need for or possible uses of theory.
\citeN{hempel-theor-dilem}, \emph{e}.\emph{g}., went so far as to
question why theories are necessary at all, given that (as he claimed)
learning at the theoretical level always follows learning at the
observational one.\footnote{\citeN{stein-newt-st} provides a
  compelling confutation of this claim, by showing that, in Newton's
  development of his abstract framework of dynamics, his metaphysics
  of space, time and motion, and his theory of universal gravitation,
  much conceptual and theoretical clarification---learning---was
  required before the empirical data could be properly comprehended at
  all, much less used as the basis for the construction of a fully
  fledged theory.}  More recent philosophers, such as
\citeN{cartwright99}, have also wanted to do away with theories,
focusing rather on models of various sorts as the fundamental tool or
unit of analysis in science based on the idea that the world is too
complex and multifarious to be adequately captured by monolithic
theories.  Newtonian abduction provides an explicit picture that shows
why the motivations behind such questions, arguments and conclusions
are misguided.  In the context of the framework, the theory and the
concrete model enter into the explanatory relations on an even footing
(the biconditional), each providing grounds for improved and deepened
understanding of the other.  Learning about one happens simultaneously
with learning about the other, each informing and informed by the
other, inextricably linked.

Much of my discussion of confirmation implicitly rejects an idea that
seems to be popular today: that confirmation must be conceived of as
giving a reason to think the thing confirmed is true.  I, however, do
not see why one needs to be a realist in order to talk about and
accept the idea (and practice) of confirmation.  For me, if one can
show that a theoretical structure in conjunction with methodological
and epistemological principles, appropriately applied, all conduce, in
ineliminable ways, to success in the scientific enterprise, then one
has \emph{eo ipso} endowed them with some measure of confirmatory
support.  If one then wants to go on to talk about truth, one is
welcome to do so, but it is not necessary.  In any event, I take it
that asserting a \emph{theory} to be ``true'' or ``approximately
true'' differs from asserting that the theory's \emph{predictions
  about} and \emph{descriptions of} systems in the world are true or
approximately true.  One who does not endorse any of the standard
forms of scientific realism will avoid the former but can welcome the
latter, but it is only the latter that has any relevance to
confirmation.

This rejection explains my claim, expressed in \S4, that confirmation
by Newtonian abduction, though it lends confirmatory support to
Lagrangian mechanics, does not give us a reason to believe that
Lagrangian mechanics is true, and indeed that I am not even sure what
a truth claim about Lagrangian mechanics could amount to.  It is,
moreover, not only Lagrangian mechanics that I do not understand bald
truth claims about---the same goes for Newtonian mechanics, and for
every other framework.  Newtonian mechanics is not about any
particular kind of physical system, the way that Newtonian
gravitational theory is about physical systems with inertial mass,
gravitational mass, and appropriate quantities representing spatial
position and correlative spatial velocity.  I therefore think there is
no inconsistency in my being willing to countenance the idea that
Newtonian gravitational theory is true (or not), in so far as
Newtonian gravitational theory is demonstrably ``about'' a well
defined class of physical systems in the world, even though I am not
comfortable with truth claims about Newtonian mechanics itself.
Perhaps one would want to say that, in so far as (\emph{ex hypothesi})
all physical systems are treatable by theories formulated in the
context of Newtonian mechanics and so manifest the structure
characteristic of Newtonian theories inherited from the abstract
structures of the framework (algebraic structures on the family of
dynamical solutions, \emph{etc}.), the framework itself is ``true of
the entire world''.  (In that same sense, Lagrangian mechanics could
also be said to be true or false.)  I would not necessarily be opposed
to that.  I would ask only that one be clear that the relation of
framework to world is radically different than that of theory to
world, and so whatever one means in each case by its truth will not
necessarily translate straightforwardly to the other.  On a related
note, some may want to claim that what I argue may be better
understood as showing that frameworks are worthy of pursuit (or of
being ``accepted''), rather than showing that they are confirmed.  As
a good card-carrying neo-Carnapian, I reject the distinction, at least
as posed in such a flat-footed way.

Before concluding, I want to make a few remarks about the possible
disconfirmation of frameworks.  Many of those who advocate against the
idea that frameworks can be confirmed, such as Kuhn, also deny that
frameworks can be disconfirmed in principled ways.  My explication of
Newtonian abduction shows this to be wrong.  If one has a framework
that has been successful in abducting theories to treat given
phenomena, and one then shows by the construction of more finely
detailed and accurate concrete models that the phenomena in question
do not in fact manifest the structure of the framework, then one has
disconfirmed the framework, at least with regard to the given
phenomena.  If one discovers new phenomena that do not manifest the
framework's structure from the start, then one has disconfirmed the
framework's possible universal validity.  Newtonian mechanics, for
instance, would be disconfirmed by the discovery of physical systems
whose dynamical evolution can be predicted only if one includes
derivatives of spatial position of higher order than the first in the
initial data.  This is incompatible with the structure of Newton's
Second Law, which is inconsistent with such initial data.  Quantum
field theory in its current incarnations (\emph{i}.\emph{e}., anything
remotely like the framework characterized by the Wightman axioms)
would be disconfirmed by discovering a violation of causality in the
form of the observation of superluminal propagation of quantum stuff,
or by discovering phenomena that violate the Heisenberg Uncertainty
Principle.  \emph{Pace} Popper, if a theoretical structure can be
disconfirmed in such ways, it can be confirmed in the converse ways:
the more phenomena we discover satisfying the Heisenberg Uncertainty
Principle, the greater credence we should have in the soundness of the
framework of quantum field theory, at least for phenomena in its
regime of applicability.

A piece of reasoning that succeeds in applying theoretical apparatus,
based on empirical evidence, to the fruitful representation of the
world in a way that gives one warrant to accept the theoretical
apparatus and continue to use it \emph{eo ipso} provides the apparatus
with confirmatory support.  That is what confirmation is.  Newtonian
abduction does this.



\bibtocsec



\begin{thebibliography}{}

\bibitem[\protect\citeauthoryear{{Abbott, B. \emph{et al}.{\space}(LIGO
  Scientific Collaboration and Virgo Collaboration)}}{{Abbott, B. \emph{et
  al}.{\space}(LIGO Scientific Collaboration and Virgo
  Collaboration)}}{2016}]{abbott-et-obs-gw-bin-bh}
{Abbott, B. \emph{et al}.{\space}(LIGO Scientific Collaboration and Virgo
  Collaboration)} (2016, February).
\newblock Observation of gravitational waves from a binary black hole merger.
\newblock {\em Physical Review Letters\/}~{\em 116\/}(6), 061102.
\newblock doi:\href{http://dx.doi.org/10.1103/PhysRevLett.116.061102}
  {10.1103/PhysRevLett.116.061102}. Preprint:
  \href{https://arxiv.org/abs/1602.03837} {arXiv:1602.03837 [gr-qc]}.

\bibitem[\protect\citeauthoryear{Bogen and Woodward}{Bogen and
  Woodward}{1988}]{bogen-woodward-sav-pha}
Bogen, J. and J.~Woodward (1988, July).
\newblock Saving the phenomena.
\newblock {\em The Philosophical Review\/}~{\em \textsc{xcvii}\/}(3), 303--352.
\newblock doi:\href{http://dx.doi.org/10.2307/2185445} {10.2307/2185445}.

\bibitem[\protect\citeauthoryear{Brown}{Brown}{1904}]{brown-deg-acc-new-lun-thry}
Brown, E. (1904, April).
\newblock On the degree of accuracy of the new lunar theory and on the final
  values of the mean motions of the perigee and node.
\newblock {\em Monthly Notices of the Royal Astronomical Society\/}~{\em
  \textsc{lxiv}\/}(6), 524--534.
\newblock doi:\href{http://dx.doi.org/10.1093/mnras/64.6.524}
  {10.1093/mnras/64.6.524}.

\bibitem[\protect\citeauthoryear{Carnap}{Carnap}{1936}]{carnap-test-mean-1}
Carnap, R. (1936, Oct.).
\newblock Testability and meaning. {\textsc{i}}.
\newblock {\em Philosophy of Science\/}~{\em 3\/}(4), 419--471.
\newblock Part 1 of a two-part article; reprinted, with omissions, in H. Feigl
  and M. Brodbeck, eds., \emph{Readings in the Philosophy of Science}, 1953
  (New York: Appleton-Century-Crofts, Inc.), pp.~47--92.

\bibitem[\protect\citeauthoryear{Carnap}{Carnap}{1956}]{carnap-eso}
Carnap, R. (1956).
\newblock Empiricism, semantics and ontology.
\newblock In {\em Meaning and Necessity: A Study in Semantics and Modal
  Logic\/} (Second ed.)., pp.\  205--221. Chicago: The University of Chicago
  Press.
\newblock An earlier version was published in \emph{Revue Internationale de
  Philosophie} 4(1950):20--40.

\bibitem[\protect\citeauthoryear{Cartwright}{Cartwright}{1999}]{cartwright99}
Cartwright, N. (1999).
\newblock {\em The Dappled World: A Study of the Boundaries of Science}.
\newblock Cambridge: Cambridge University Press.

\bibitem[\protect\citeauthoryear{Curiel}{Curiel}{2001}]{curiel-modesty}
Curiel, E. (2001).
\newblock A plea for modesty: {A}gainst the current excesses in quantum
  gravity.
\newblock {\em Philosophy of Science\/}~{\em 68\/}(3), S424--S441.
\newblock doi:\href{http://dx.doi.org/10.1086/392926} {10.1086/392926}.

\bibitem[\protect\citeauthoryear{Curiel}{Curiel}{2014}]{curiel-cm-lag-not-ham}
Curiel, E. (2014).
\newblock Classical mechanics is {L}agrangian; it is not {H}amiltonian.
\newblock {\em British Journal for the Philosophy of Science\/}~{\em 65\/}(2),
  269--321.
\newblock doi:\href{http://dx.doi.org/10.1093/bjps/axs034}
  {10.1093/bjps/axs034}.

\bibitem[\protect\citeauthoryear{Curiel}{Curiel}{2017a}]{curiel-kins-dyns-struc-theors}
Curiel, E. (2017a).
\newblock Kinematics, dynamics, and the structure of physical theory.
\newblock Unpublished manuscript. Preprint:
\href{http://arxiv.org/abs/1603.02999} {arXiv:1603.02999 [gr-qc]}.  Most recent draft available at
  $<$\url{http://strangebeautiful.com/papers/curiel-kins-dyns-struc-theory.pdf}$>$.

\bibitem[\protect\citeauthoryear{Curiel}{Curiel}{2017b}]{curiel-meas-topo-prob-cosmo}
Curiel, E. (2017b).
\newblock Measure, topology and probabilistic reasoning in cosmology.
\newblock Preprint: \href{http://arxiv.org/abs/1509.01878} {arXiv:1509.01878
  [gr-qc]}.  Most recent draft available at
  $<$\url{http://strangebeautiful.com/papers/curiel-meas-topo-prob-cosmo.pdf}$>$.

\bibitem[\protect\citeauthoryear{Curiel}{Curiel}{2017c}]{curiel-prpy-basis-sems}
Curiel, E. (2017c).
\newblock On the propriety of physical theories as a basis for their semantics.
\newblock Unpublished manuscript.
  Most recent draft available at
  $<$\url{http://strangebeautiful.com/papers/curiel-propriety-adequacy-basis-sems.pdf}$>$.

\bibitem[\protect\citeauthoryear{Curiel}{Curiel}{2019}]{curiel-schem-obsr-epi-cont-theors}
Curiel, E. (2019).
\newblock Schematizing the observer and the epistemic content of theories.
\newblock Forthcoming in \emph{Studies in History and Philosophy of Modern
  Physics}. Preprint: \href{https://arxiv.org/abs/1903.02182} {arXiv:1903.02182
  [physics.hist-ph]}.

\bibitem[\protect\citeauthoryear{Earman and Salmon}{Earman and
  Salmon}{1992}]{earman-salmon-conf-sci-hypos}
Earman, J. and W.~Salmon (1992).
\newblock The confirmation of scientific hypotheses.
\newblock In M.~Salmon, J.~Earman, C.~Glymour, and J.~Lennox (Eds.), {\em
  Introduction to the Philosophy of Science}, pp.\  42--103. Englewood Cliff,
  NJ: Prentice Hall.

\bibitem[\protect\citeauthoryear{Fermi}{Fermi}{1937}]{fermi-thermo}
Fermi, E. (1937).
\newblock {\em Thermodynamics}.
\newblock New York: Dover Publications, Inc.
\newblock The 1956 Dover edition is an unabridged, unaltered republication of
  the 1937 Prentice-Hall edition.

\bibitem[\protect\citeauthoryear{Fowler and Guggenheim}{Fowler and
  Guggenheim}{1949}]{fowler-guggenheim-stat-thermo-2nd}
Fowler, R. and E.~Guggenheim (1949).
\newblock {\em Statistical Thermodynamics: {A} Version of Statistical Mechanics
  for Students of Physics and Chemistry\/} (Second ed.).
\newblock Cambridge: Cambridge University Press.

\bibitem[\protect\citeauthoryear{Freundlich}{Freundlich}{1915}]{freundlich-planet}
Freundlich, E. (1915).
\newblock {\"U}ber die {E}rkl\"arung der {A}nomalien im {P}laneten-{S}ystem
  durch die {G}ravitationswirkung interplanetarer {M}assen.
\newblock {\em Astronomische Nachrichten\/}~{\em 201\/}(4803),
49--56. doi:\href{http://dx.doi.org/10.1002/asna.19152010302} {10.1002/asna.19152010302}

\bibitem[\protect\citeauthoryear{Friedman}{Friedman}{2001}]{friedman-dyns-reason-addend}
Friedman, M. (2001).
\newblock {\em The Dynamics of Reason}.
\newblock Stanford, CA: CSLI Publications.

\bibitem[\protect\citeauthoryear{Frieman, Turner, and Huterer}{Frieman
  et~al.}{2008}]{frieman-et-dark-energy-acc-uni}
Frieman, J., M.~Turner, and D.~Huterer (2008, September).
\newblock Dark energy and the accelerating universe.
\newblock {\em Annual Review of Astronomy and Astrophysics\/}~{\em 46},
  385--432.
\newblock doi:\href{http://dx.doi.org/10.1146/annurev.astro.46.060407.145243}
  {10.1146/annurev.astro.46.060407.145243}. Preprint:
  \href{https://arxiv.org/abs/0803.0982} {arXiv:0803.0982 [astro-ph]}.

\bibitem[\protect\citeauthoryear{Griffiths and Podolsk\'y}{Griffiths and
  Podolsk\'y}{2009}]{griffiths-podolsky-exact-sts}
Griffiths, J. and J.~Podolsk\'y (2009).
\newblock {\em Exact Space-Times in Einstein's General Relativity}.
\newblock Cambridge: Cambridge University Press.

\bibitem[\protect\citeauthoryear{Hall}{Hall}{1894}]{hall-sugg-theory-merc}
Hall, A. (1894, June).
\newblock A suggestion in the theory of mercury.
\newblock {\em The Astronomical Journal\/}~{\em \textsc{xiv}\/}(319), 49--51.
\newblock doi:\href{http://dx.doi.org/10.1086/102055} {10.1086/102055}.

\bibitem[\protect\citeauthoryear{Harper}{Harper}{1990}]{harper-newt-class-deduc-pha}
Harper, W. (1990).
\newblock Newton's classic deductions from phenomena.
\newblock {\em PSA: Proceedings of the Biennial Meeting of the Philosophy of
  Science Association\/}~{\em 2}, 183--196.
\newblock Stable URL: $<$\url{https://www.jstor.org/stable/193067}$>$.

\bibitem[\protect\citeauthoryear{Harper}{Harper}{1999}]{harper-1st-6-props-newt-arg}
Harper, W. (1999).
\newblock The first six propositions in {N}ewton's argument for universal
  gravity.
\newblock {\em The St.\@ John's Review\/}~{\em \textsc{xlv}\/}(2), 74--93.
\newblock In the special issue of the proceedings of the conference ``Beyond
  Hypothesis: Newton's Experimental Philosophy'', St. John's College,
  Annapolis, MD 19--21 March, 1999. Electronic copy available at
  $<$\url{http://digitalarchives.sjc.edu/files/original/5d140db01fbf5fe6562615e2d3713b6e.pdf}$>$.

\bibitem[\protect\citeauthoryear{Harper}{Harper}{2011}]{harper-newtons-sci-meth}
Harper, W. (2011).
\newblock {\em Isaac Newton's Scientific Method: Turning Data into Evidence
  about Gravity and Cosmology}.
\newblock Oxford: Oxford University Press.

\bibitem[\protect\citeauthoryear{Hempel}{Hempel}{1955}]{hempel-funds-conc-form}
Hempel, C. (1955).
\newblock Fundamentals of concept formation in empirical science.
\newblock In O.~Neurath, R.~Carnap, and C.~Morris (Eds.), {\em Foundations of
  the Unity of Science: Toward an International Encyclopedia of Unified
  Science}, Volume \textsc{ii}, Chapter~7. Chicago: The University of Chicago
  Press.
\newblock Formerly entitled \emph{International Encyclopedia of Unified
  Science}.

\bibitem[\protect\citeauthoryear{Hempel}{Hempel}{1965}]{hempel-theor-dilem}
Hempel, C. (1965).
\newblock The theoretician's dilemma: {A} study in the logic of theory
  construction.
\newblock In {\em Aspects of Scientific Explanation: {A}nd Other Essays in the
  Philosophy of Science}, Chapter~8, pp.\  173--226. New York: Free Press.
\newblock Originally published in H. Feigl, \emph{et al}.\ (eds.),
  \emph{Concepts, Theories and the Mind-Body Problem}, Minnesota Studies in
  Philosophy of Science Vol.~\textsc{ii}, Minneapolis: University of Minnesota
  Press, 1958, pp.~37--98.

\bibitem[\protect\citeauthoryear{Henderson}{Henderson}{2014}]{henderson-bayes-ibe}
Henderson, L. (2014, December).
\newblock Bayesianism and inference to the best explanation.
\newblock {\em British Journal for the Philosophy of Science\/}~{\em 65\/}(4),
  687--715.
\newblock doi:\href{http://dx.doi.org/10.1093/bjps/axt020}
  {10.1093/bjps/axt020}.

\bibitem[\protect\citeauthoryear{Henderson, Goodman, Tenenbaum, and
  Woodward}{Henderson et~al.}{2010}]{henderson-et-sci-thrs-hier-bayes}
Henderson, L., N.~Goodman, J.~Tenenbaum, and J.~Woodward (2010, April).
\newblock The structure and dynamics of scientific theories: {A} hierarchical
  {B}ayesian perspective.
\newblock {\em Philosophy of Science\/}~{\em 77\/}(2), 172--200.
\newblock doi:\href{http://dx.doi.org/10.1086/651319} {10.1086/651319}.

\bibitem[\protect\citeauthoryear{Hertz}{Hertz}{1884}]{hertz-bezich-maxwell-gegner}
Hertz, H. (1884).
\newblock \"Uber die {B}ezichungen zwischen den {M}axwell'schen
  electrodynamischen {G}rundgleichungen und den {G}rundgleichungen der
  gegnerischen {E}lectrodynamik.
\newblock {\em Annalen der Physik und Chemie\/}~{\em \textsc{xxiii}}, 84--103.
\newblock Translated D. E. Jones and G. A. Schott as ``On the Relations between
  Maxwell's Fundamental Electromagnetic Equations and the Fundamental Equations
  of the Opposing Electromagnetics'', in H. Hertz \emph{Miscellaneous Papers},
  London:MacMillan and Co., Ltd., ch.~\textsc{xvii}, pp.~273--290.
\newblock doi:\href{http://dx.doi.org/10.1002/andp.18842590904} {10.1002/andp.18842590904}.

\bibitem[\protect\citeauthoryear{Kuhn}{Kuhn}{1996}]{kuhn-struc-sci-rev}
Kuhn, T. (1996).
\newblock {\em The Structure of Scientific Revolutions\/} (Third edition ed.).
\newblock Chicago: University of Chicago Press.
\newblock The first edition was published in 1962.

\bibitem[\protect\citeauthoryear{Kuipers}{Kuipers}{2001}]{kuipers-strucs-sci-heur-patts}
Kuipers, T. (2001).
\newblock {\em Structures in Science: {H}euristic Patterns Based on Cognitive
  Structures; {A}n Advanced Textbook in Neo-Classical Philosophy of Science},
  Volume 301 of {\em Synthese Library}.
\newblock Dordrecht: Kluwer.

\bibitem[\protect\citeauthoryear{Lakatos}{Lakatos}{1970}]{lakatos-fals-meth-sci-rsrch}
Lakatos, I. (1970).
\newblock Falsification and the methodology of scientific research programmes.
\newblock In I.~Lakatos and A.~Musgrave (Eds.), {\em Criticism and the Growth
  of Knowledge}, pp.\  91--196. Cambridge: Cambridge University Press.
\newblock Reprinted in his \emph{Methodology of Scientific Research Programmes:
  Philosophical Papers}, vol.~1, Cambridge: Cambridge University Press, 1980,
  eds.~J. Worrall and G. Currie, ch.~1, pp.~8--101.

\bibitem[\protect\citeauthoryear{Landau and Lifschitz}{Landau and
  Lifschitz}{1975}]{landau-lifschitz-fluid}
Landau, L. and E.~Lifschitz (1975).
\newblock {\em Fluid Mechanics\/} (Second ed.).
\newblock Oxford: Pergamon Press.
\newblock An expanded, revised edition of the original 1959 edition. Translated
  from the Russian by J. Sykes and W. Reid.

\bibitem[\protect\citeauthoryear{Landau and Lifschitz}{Landau and
  Lifschitz}{1980}]{landau-lifschitz-stat-phys-1}
Landau, L. and E.~Lifschitz (1980).
\newblock {\em Statistical Physics, Part 1\/} (Third ed.).
\newblock Oxford: Pergamon Press.
\newblock An expanded, revised edition of the original 1959 edition, by E.
  Lifschitz and L. Pitaevskii. Translated from the the Russian by J. Sykes and
  W. Reid.

\bibitem[\protect\citeauthoryear{Laudan}{Laudan}{1977}]{laudan-prog-probs}
Laudan, L. (1977).
\newblock {\em Progress and Its Problems: Towards a Theory of Scientific
  Growth}.
\newblock Berkeley: University of California Press.

\bibitem[\protect\citeauthoryear{Le~Verrier}{Le~Verrier}{1845}]{leverrier-mem-vars-orbs}
Le~Verrier, U. (1845).
\newblock {\em M\'emoire sur les Variations S\'eculaires des \'El\'ements des
  Orbites, pour les Sept Plan\`etes Principales, Mercure, Venus, la Terre,
  Mars, Jupiter, Saturne, Uranus}.
\newblock Paris: Bachelier.

\bibitem[\protect\citeauthoryear{Le~Verrier}{Le~Verrier}{1859}]{leverrier-merc}
Le~Verrier, U. (1859).
\newblock Lettre de {M}. {L}e {V}errier \`a {M}. {F}aye sur la th\'eorie de
  {M}ercure et sur le mouvement du p\'erih\'elie de cette plan\`ete.
\newblock {\em Comptes rendus hebdomadaires des s\'eances de l'Acad\'emie des
  sciences (Paris)\/}~{\em 49}, 379--383.

\bibitem[\protect\citeauthoryear{Lipton}{Lipton}{2004}]{lipton-ibe}
Lipton, P. (2004).
\newblock {\em Inference to the Best Explanation\/} (Second ed.).
\newblock London: Routledge.

\bibitem[\protect\citeauthoryear{Malament}{Malament}{1986}]{malament-newt-grav-geom-spc}
Malament, D. (1986).
\newblock {N}ewtonian gravity, limits, and the geometry of space.
\newblock In R.~Colodny (Ed.), {\em From Quarks to Quasars: Philosophical
  Problems of Modern Physics}, pp.\  181--201. Pittsburgh: Pittsburgh
  University Press.

\bibitem[\protect\citeauthoryear{Malament and Zabell}{Malament and
  Zabell}{1980}]{malament-zabell-gibbs-phase-avg}
Malament, D. and S.~Zabell (1980, September).
\newblock Why {G}ibbs phase averages work---{T}he role of ergodic theory.
\newblock {\em Philosophy of Science\/}~{\em 47\/}(3), 339--349.
\newblock doi:\href{http://dx.doi.org/10.1086/288941} {10.1086/288941}.

\bibitem[\protect\citeauthoryear{Maxwell}{Maxwell}{1856}]{maxwell-farad-line}
Maxwell, J.~C. (1856).
\newblock On {F}araday's lines of force.
\newblock See \citeN{maxwell-coll-paps}, pp.\  155--229.
\newblock Originally read before the Cambridge Philosophical Society on
  December 10, 1855, and February 11, 1856, and subsequently published in the
  \emph{Transactions of the Cambridge Philosophical Society}, 1864, \textsc{x},
  part~\textsc{i}, 27--83.

\bibitem[\protect\citeauthoryear{Maxwell}{Maxwell}{1864}]{maxwell-dyn-emf}
Maxwell, J.~C. (1864).
\newblock A dynamical theory of the electromagnetic field.
\newblock See \citeN{maxwell-coll-paps}, pp.\  526--597.
\newblock Originally published in three parts: Part \textsc{i}, read before the
  Royal Society on December 8, 1864, and subsequently published in the
  \emph{Royal Society Proceedings}, \textsc{iii}, 531--536. Part \textsc{ii}:
  \emph{Philosophical Transactions of the Royal Society}, 1865, \textsc{clv},
  459--512. Part \textsc{iii}: \emph{Philosophical Magazine}, 1865,
  \textsc{xxix}, 152--157.

\bibitem[\protect\citeauthoryear{Maxwell}{Maxwell}{1876}]{maxwell-eom-conn-sys-in65}
Maxwell, J.~C. (1876).
\newblock On the proof of the equations of motion of a connected system.
\newblock {\em Proceedings of the Cambridge Philosophical Society\/}~{\em
  \textsc{ii}}, 292--294.
\newblock Reprinted in Maxwell 1965, volume \textsc{ii}, 308--309.

\bibitem[\protect\citeauthoryear{Maxwell}{Maxwell}{1891}]{maxwell-treat-em}
Maxwell, J.~C. (1891).
\newblock {\em A Treatise on Electricity and Magnetism\/} (3rd ed.).
\newblock New York: Dover Publications, Inc.
\newblock In 2 volumes. 1954 facsimile of the edition originally published by
  Clarendon Press in 1891, edited by J.~J.~Thomson.

\bibitem[\protect\citeauthoryear{Maxwell}{Maxwell}{1965b}]{maxwell-coll-paps}
Maxwell, J.~C. (1965b).
\newblock {\em The Scientific Papers of {J.~C.~Maxwell}}.
\newblock New York: Dover Publications, Inc.
\newblock W.~Niven (Ed.). Two volumes, published as one.

\bibitem[\protect\citeauthoryear{Newcomb}{Newcomb}{1895a}]{newcomb-4-inner-planets}
Newcomb, S. (1895a).
\newblock {\em The Elements of the Four Inner Planets and the Fundamental
  Constants of Astronomy}.
\newblock Washington, DC: US Government Printing Office.
\newblock Supplement to the American Ephemeris and Nautical Almanac for 1897.

\bibitem[\protect\citeauthoryear{Newcomb}{Newcomb}{1895b}]{newcomb-fund-consts-astro}
Newcomb, S. (1895b).
\newblock On the principal fundamental constants of astronomy.
\newblock {\em Astronomical Journal\/}~{\em 14}, 185--189.
\newblock doi:\href{http://dx.doi.org/10.1086/102176} {10.1086/102176.}

\bibitem[\protect\citeauthoryear{Newcomb}{Newcomb}{1905}]{newcomb-precess-consts}
Newcomb, S. (1905).
\newblock {\em A New Determination of the Precessional Constant with the
  Resulting Precessional Motions}.
\newblock Washington, DC: U.S. Nautical Almanac Office.

\bibitem[\protect\citeauthoryear{Newcomb}{Newcomb}{1911}]{newcomb-grav-encyc-brit}
Newcomb, S. (1911).
\newblock Gravitation.
\newblock In H.~Chisholm (Ed.), {\em The Encyclop{\ae}dia Britannica: A
  Dictionary of Arts, Sciences, Literature and General Information\/} (Eleventh
  edition ed.), Volume \textsc{xii}, pp.\  384--389. Cambridge: University of
  Cambridge Press.

\bibitem[\protect\citeauthoryear{Newton}{Newton}{1672}]{newton-light-color}
Newton, I. (1672).
\newblock Letter of february 6, 1671/72, to {H}enry {O}ldenburg, {S}ecretary of
  the {R}oyal {S}ociety, outlining {N}ewton's researches on light and color.
\newblock In I.~Cohen (Ed.), {\em {Isaac Newton's} Papers \& Letters on Natural
  Philosophy}, pp.\  47--59. Cambridge, MA: Harvard University Press.
\newblock The original was published in the \emph{Philosophical Transactions of
  the Royal Society}, 80(February 16, 1671/72):3075--3087.

\bibitem[\protect\citeauthoryear{Newton}{Newton}{1726}]{newton-princ-motte}
Newton, I. (1726).
\newblock {\em Philosophi{\ae} Naturalis Principia Mathematica\/} (Third ed.).
\newblock Amherst, NY: Prometheus Press.
\newblock The translation by A.~Motte of the third edition (1726), originally
  produced in 1729, and reprinted by Prometheus Press in 1995. The first
  edition of the \emph{Principia} was published in 1686, the second in 1713.

\bibitem[\protect\citeauthoryear{Norton}{Norton}{2005}]{norton-little-survey-induc}
Norton, J. (2005).
\newblock A little survey on induction.
\newblock In P.~Achinstein (Ed.), {\em Scientific Evidence: {P}hilosophical
  Theories and Applications}, pp.\  9--34. Baltimore: John Hopkins University
  Press.

\bibitem[\protect\citeauthoryear{Peirce}{Peirce}{1878a}]{peirce-ded-ind-hypo}
Peirce, C.~S. (1878a).
\newblock Deduction, induction and hypothesis.
\newblock See \citeN{peirce-sel-phil-wrtg-i}, Chapter~12, pp.\  186--199.
\newblock Originally published in \emph{Popular Science Monthly} 13(August,
  1878):470--482.

\bibitem[\protect\citeauthoryear{Peirce}{Peirce}{1878b}]{peirce-doct-chnc}
Peirce, C.~S. (1878b).
\newblock The doctrine of chances.
\newblock See \citeN{peirce-sel-phil-wrtg-i}, Chapter~9, pp.\  142--154.
\newblock Originally published in \emph{Popular Science Monthly} 12(March,
  1878):604--615.

\bibitem[\protect\citeauthoryear{Peirce}{Peirce}{1903}]{peirce-nat-mean}
Peirce, C.~S. (1903).
\newblock The nature of meaning.
\newblock In {\em The Essential Peirce: Selected Philosophical Writings},
  Volume 2 (1893--1913), Chapter~15, pp.\  208--225. Bloomington, IN: Indiana
  University Press.
\newblock Originally delivered 07 May 1903 at Harvard University, as the sixth
  in a series of seven public lectures on pragmatism.

\bibitem[\protect\citeauthoryear{Peirce}{Peirce}{1955}]{peirce-abd-ind}
Peirce, C.~S. (1955).
\newblock Abduction and induction.
\newblock In {\em Philosophical Writings of Peirce}, Chapter~11, pp.\
  150--156. New York: Dover Publications, Inc.
\newblock Collated by the editor from 4 different manuscripts, 3 of them
  unpublished.

\bibitem[\protect\citeauthoryear{Peirce}{Peirce}{1992}]{peirce-sel-phil-wrtg-i}
Peirce, C.~S. (1992).
\newblock {\em The Essential Peirce: Selected Philosophical Writings}, Volume 1
  (1867--1893).
\newblock Bloomington, IN: Indiana University Press.
\newblock Edited by N. Houser and C. Kloesel.

\bibitem[\protect\citeauthoryear{Romeijn}{Romeijn}{2013}]{romeijn-abd-bayes}
Romeijn, J.-W. (2013, December).
\newblock Abducted by bayesians?
\newblock {\em Journal of Applied Logic\/}~{\em 11\/}(4), 430--439.
\newblock Part of special issue ``Combining Probability and Logic: Papers from
  Progic 2011'', ed.\@ J. Helzner.
  doi:\href{http://dx.doi.org/10.1016/j.jal.2012.09.003}
  {10.1016/j.jal.2012.09.003}.

\bibitem[\protect\citeauthoryear{Sakurai}{Sakurai}{1964}]{sakurai-invar-princs-elem-parts}
Sakurai, J. (1964).
\newblock {\em Invariance Principles and Elementary Particles}.
\newblock Princeton: Princeton University Press.

\bibitem[\protect\citeauthoryear{Smith}{Smith}{2014}]{smith-closing-loop}
Smith, G. (2014).
\newblock Closing the loop: {T}esting {N}ewtonian gravity then and now.
\newblock In Z.~Biener and E.~Schliesser (Eds.), {\em Newton and Empiricism},
  Chapter~10, pp.\  262--351. Oxford: Oxford University Press.
\newblock doi:\href{http://dx.doi.org/10.1093/acprof:oso/9780199337095.003.0011}
  {10.1093/acprof:oso/9780199337095.003.0011}.

\bibitem[\protect\citeauthoryear{Sommerfeld}{Sommerfeld}{1964}]{sommerfeld-thermo}
Sommerfeld, A. (1964).
\newblock {\em Thermodynamics and Statistical Mechanics}, Volume \textsc{v} of
  {\em Lectures on Theoretical Physics}.
\newblock New York: Academic Press.
\newblock Trans. J. Kestin. Edited and posthumously completed by F. Bopp and J.
  Meixner.

\bibitem[\protect\citeauthoryear{Sorkin}{Sorkin}{1979}]{sorkin-mean-canon-ensemb}
Sorkin, R. (1979, May).
\newblock On the meaning of the canonical ensemble.
\newblock {\em International Journal of Theoretical Physics\/}~{\em 18\/}(5),
  309--321.
\newblock doi:\href{http://dx.doi.org/10.1007/BF00670427} {10.1007/BF00670427}.

\bibitem[\protect\citeauthoryear{Stanford}{Stanford}{2006}]{stanford-unconc-alters}
Stanford, P. (2006).
\newblock {\em Exceeding Our Grasp: {S}cience, History, and the Problem of
  Unconceived Alternatives}.
\newblock Oxford: Oxford University Press.
\newblock doi:\href{http://dx.doi.org/10.1093/0195174089.001.0001}
  {10.1093/0195174089.001.0001}.

\bibitem[\protect\citeauthoryear{Stein}{Stein}{1967}]{stein-newt-st}
Stein, H. (1967).
\newblock Newtonian space-time.
\newblock {\em Texas Quarterly\/}~{\em 10}, 174--200.

\bibitem[\protect\citeauthoryear{Stein}{Stein}{1989}]{stein-yes-but}
Stein, H. (1989, June).
\newblock Yes, but\ldots: Some skeptical remarks on realism and anti-realism.
\newblock {\em Dialectica\/}~{\em 43\/}(1-2), 47--65.
\newblock doi:\href{http://dx.doi.org/10.1111/j.1746-8361.1989.tb00930.x}
  {10.1111/j.1746-8361.1989.tb00930.x}.

\bibitem[\protect\citeauthoryear{Stein}{Stein}{1990}]{stein-deduct-hypo}
Stein, H. (1990).
\newblock ``{F}rom the ph{\ae}nomena of motions to the forces of nature'':
  {H}ypothesis or deduction?
\newblock {\em PSA: Proceedings of the Biennial Meeting of the
  Philosophy of Science Association\/}~{\em 2}, 209--222.  Stable URL: <\url{http://www.jstor.org/stable/193069}>.

\bibitem[\protect\citeauthoryear{Stein}{Stein}{1992}]{stein-carnap-not-wrong}
Stein, H. (1992).
\newblock Was {C}arnap entirely wrong, after all?
\newblock {\em Synthese\/}~{\em 93}, 275--295.
\newblock doi:\href{http://dx.doi.org/10.1007/BF00869429} {10.1007/BF00869429}.

\bibitem[\protect\citeauthoryear{Stein}{Stein}{1994}]{stein-struct-know}
Stein, H. (1994).
\newblock Some reflections on the structure of our knowledge in physics.
\newblock In D.~Prawitz, B.~Skyrms, and D.~Westerst{\aa}hl (Eds.), {\em Logic,
  Metholodogy and Philosophy of Science}, Proceedings of the Ninth
  International Congress of Logic, Methodology and Philosophy of Science, pp.\
  633--655. New York: Elsevier Science B.V.

\bibitem[\protect\citeauthoryear{Stein}{Stein}{unpublished-a}]{stein-furth-consid-newt-meth}
Stein, H. (unpublished-a).
\newblock Further considerations on {N}ewton's method.
\newblock Unpublished manuscript. Available at
  $<$\url{http://strangebeautiful.com/phil-phys.html}$>$.

\bibitem[\protect\citeauthoryear{Stein}{Stein}{unpublished-b}]{stein-meta-meth-newton}
Stein, H. (unpublished-b).
\newblock On metaphysics and method in {N}ewton.
\newblock Unpublished manuscript. Available at
  $<$\url{http://strangebeautiful.com/phil-phys.html}$>$.

\bibitem[\protect\citeauthoryear{Szekeres}{Szekeres}{1975}]{szekeres-inhom-cosm-mods}
Szekeres, P. (1975).
\newblock A class of inhomogeneous cosmological models.
\newblock {\em Communications in Mathematical Physics\/}~{\em 41\/}(1), 55--64.
\newblock Open-access text available at:
  \url{http://projecteuclid.org/euclid.cmp/1103860587}.
  \href{http://dx.doi.org/10.1007/BF01608547} {10.1007/BF01608547}.

\bibitem[\protect\citeauthoryear{The\u{s}i\'c, Eva, and Hartmann}{The\u{s}i\'c
  et~al.}{2017}]{tesic-et-conf-explan-bayes-ibe}
The\u{s}i\'c, M., B.~Eva, and S.~Hartmann (2017).
\newblock Confirmation by explanation: {A} {B}ayesian justification of {IBE}.
\newblock Unpublished manuscript.  Preprint: $<$\url{http://philsci-archive.pitt.edu/13328/}$>$.

\bibitem[\protect\citeauthoryear{{}van Fraassen}{{}van
  Fraassen}{1990}]{fraassen-laws-sym}
{}van Fraassen, B. (1990).
\newblock {\em Laws and Symmetry}.
\newblock Oxford: Oxford University Press.

\bibitem[\protect\citeauthoryear{Wald}{Wald}{1984}]{wald-gr}
Wald, R. (1984).
\newblock {\em General Relativity}.
\newblock Chicago: University of Chicago Press.

\bibitem[\protect\citeauthoryear{Weinberg}{Weinberg}{2008}]{weinberg-cosmo}
Weinberg, S. (2008).
\newblock {\em Cosmology}.
\newblock Oxford: Oxford University Press.

\bibitem[\protect\citeauthoryear{Wilson}{Wilson}{2004}]{wilson-newt-celest-mechs}
Wilson, C. (2004).
\newblock Newton and celestial mechanics.
\newblock In I.~Cohen and G.~Smith (Eds.), {\em The Cambridge Companion to
  Newton}, Chapter~6, pp.\  202--226. Cambridge: Cambridge University Press.

\bibitem[\protect\citeauthoryear{Worrall}{Worrall}{2000}]{worrall-scope-lims-meth-deduc-pha}
Worrall, J. (2000).
\newblock The scope, limits, and distinctiveness of the method of `deduction
  from the phenomena': {S}ome lessons from {N}ewton's `demonstrations' in
  optics.
\newblock {\em British Journal for the Philosophy of Science\/}~{\em 51},
  45--80.
\newblock \href{http://dx.doi.org/10.1093/bjps/51.1.45} {10.1093/bjps/51.1.45}.
  
\bibitem[\protect\citeauthoryear{Yunes and Pretorius}{Yunes and
  Pretorius}{2009}]{yunes-pretorius-theor-bias-grav-wv-ppe}
Yunes, N. and F.~Pretorius (2009, December).
\newblock Fundamental theoretical bias in gravitational wave astrophysics and
  the parametrized post-{E}insteinian framework.
\newblock {\em Physical Review D\/}~{\em 80\/}(12), 122003.
\newblock doi:\href{http://dx.doi.org/10.1103/PhysRevD.80.122003}
  {10.1103/PhysRevD.80.122003}.

\end{thebibliography}
\end{document}